# A Comparative Study of the Usability of Two Object-oriented Concurrent Programming Languages

Sebastian Nanz[1]   Faraz Torshizi[2]   Michela Pedroni[1]   Bertrand Meyer[1]

[1]ETH Zurich   [2]University of Toronto
firstname.lastname@inf.ethz.ch   faraz@cs.toronto.edu

**Abstract**

Concurrency has been rapidly gaining importance in general-purpose computing, caused by the recent turn towards multicore processing architectures. As a result, an increasing number of developers have to learn to write concurrent programs, a task that is known to be hard even for the expert. Language designers are therefore working on languages that promise to make concurrent programming "easier" than using traditional thread libraries. However, the claim that a new language is more usable than another cannot be supported by purely theoretical considerations, but calls for empirical studies. In this paper, we present the design of a study to compare concurrent programming languages with respect to comprehending and debugging existing programs and writing correct new programs. A critical challenge for such a study is avoiding the bias that might be introduced during the training phase and when interpreting participants' solutions. We address these issues by the use of self-study material and an evaluation scheme that exposes any subjective decisions of the corrector, or eliminates them altogether. We apply our design to a comparison of two object-oriented languages for concurrency, multithreaded Java and SCOOP (Simple Concurrent Object-Oriented Programming), in an academic setting. We obtain results in favor of SCOOP even though the study participants had previous training in Java Threads.

# Contents

# 1  Introduction

Concurrent programming has been practiced for over 40 years, but was until recently perceived as a task for specialists in high-performance computing, operating systems or networking. The transition to parallel architectures, in particular to multicore processors, has radically changed this situation, making concurrency a concern for mainstream software development.

The new situation entails that many programmers without extensive concurrency training have to write concurrent programs, a task widely acknowledged as error-prone due to concurrency-specific errors, e.g. data races or deadlocks. Such errors typically arise from incorrect use of synchronization primitives such as locks and semaphores, which traditionally are provided by concurrency libraries.

To avoid the pitfalls of the library approach, the programming languages community works towards integrating concurrency mechanisms into new languages. The goal is to raise the level of abstraction for expressing concurrency and synchronization, and hence to make programmers produce better code. Resulting programming models can exclude certain classes of errors by construction, usually accepting a penalty in performance or programming flexibility for the sake of program correctness.

The question remains whether these new languages can deliver and indeed make concurrent programming "easier" for the developer: both understanding and modification of existing code and the production of new correct code should be improved. It is difficult to argue for such properties in an abstract manner as they are connected to human subjects: empirical analyses of the usability of concurrent languages are needed to distinguish promising from less promising approaches, driving language research in the right direction.

Empirical studies for this purpose have to deal with two main challenges. First, to compare the usability of two languages side-by-side, additional programmer training is typically needed: only few programmers will be skilled in two or more novel programming paradigms. However, bias introduced during the training process has to be avoided at any cost. Second, a test to judge the proficiency of participants using the languages has to be developed, along with objective means to interpret participants' answers.

In this paper we propose the design of an empirical study that addresses the mentioned challenges and provides a template for comparing concurrent programming languages. In particular, we make the following contributions:

- a design for comparative studies of concurrent programming languages, based on self-study followed by individual tests;

- a template for a self-study document to learn the basics of concurrency and a new concurrent language;

- a set of test questions that allows for a direct comparison of approaches;

- an evaluation scheme for interpreting answers to the test questions, objective and reproducible;

- application of the study design to a comparison of two concrete languages, multithreaded Java and SCOOP, in an academic setting with 67 B.Sc. students.

The remainder of this paper is structured as follows. In Section 2 we review multithreaded Java and SCOOP and give an overview of related approaches to concurrent programming. Section 3 explores potential hypotheses of comparative studies of concurrent languages and outlines our choices. In Section 4 we present the an overview of the design of the study. We present the design of the training phase including the structure for a self-study document on concurrency in Section 5. The design of the test and the results of the multithreaded Java vs. SCOOP study are presented in Section 6. We discuss threats to validity in Section 7 and give an overview of related work in Section 8. We conclude and present avenues for future work in Section 9.

In this extended report, the complete self-study material and the tests, as used in the concrete multithreaded Java vs. SCOOP study, is found in Appendix A–D.

## 2 Concurrent programming languages

As background for the main part of the paper, this section briefly reviews SCOOP (Simple Concurrent Object-Oriented Programming) [14, 16] and multithreaded Java [20], two object-oriented concurrent programming models. The section concludes with an overview of related approaches.

### 2.1 SCOOP

The central idea of SCOOP is that every object is associated for its lifetime with a *processor*, an abstract notion denoting a site for computation: just as threads may be assigned to cores on a multicore system, processors may be assigned to cores, or even to remote processing units. References can point to local objects (on the same processor) or to objects on other processors; the latter ones are called *separate* references. Calls within a single processor remain synchronous, while calls to objects on other processors are dispatched asynchronously to those processors for execution, thus giving rise to concurrent execution.

The SCOOP version of the producer/consumer problem serves as a simple illustration of these main ideas. In a root class, the main entities *producer* and *consumer* are defined. The keyword **separate** denotes that these entities may be associated with a processor different from the current one.

    *producer*: **separate** *PRODUCER*
    *consumer*: **separate** *CONSUMER*

Creation of an separate object such as *producer* results in the creation of a new processor and of a new object of type *PRODUCER* that is associated with this processor. Hence in this example, calls to *producer* and *consumer* will be executed concurrently, as they will be associated with two different new processors.

Both *producer* and *consumer* access an unbounded buffer

    *buffer*: **separate** *BUFFER* [**INTEGER**]

and thus their access attempts need to be synchronized to avoid data races (by *mutual exclusion*) and to avoid that an empty buffer is accessed (by *condition synchronization*). To ensure mutual exclusion, processors that are needed for the execution of a routine are automatically locked by the runtime system before entering the body of the routine. The model prescribes that separate objects needed by are routine are *controlled*, i.e. passed as arguments to the routine.

For example, in a call *consume*(*buffer*), the separate object *buffer* is controlled and thus the processor associated with *buffer* gets locked. This prevents data races on this object for the duration of the routine. For condition synchronization, the condition to be waited upon can be explicitly stated as a precondition, indicated by the keyword **require**. The evaluation of the condition uses wait semantics: the runtime system automatically delays the routine execution until the condition is true. For example, the implementation of the routine *consume*, defined in the consumer, ensures that an item from *a_buffer* is only removed if *a_buffer* is not empty:

```
consume (a_buffer: separate BUFFER[INTEGER])
    require
        not (a_buffer.count = 0)
    local
        value: INTEGER
    do
        value := a_buffer.get
    end
```

Note that the runtime system further ensures that the result of the call *a_buffer.get* is properly assigned to *value* using a mechanism called *wait by necessity*: while the client usually does not have to wait for an asynchronous call to finish, it will do so if it needs the result of this call.

The corresponding producer routine does not need a condition to be waited upon (unboundedness of the buffer):

```
produce (a_buffer: separate BUFFER[INTEGER])
    local
        value: INTEGER
    do
        value := new_value
        a_buffer.put (value)
    end
```

In summary, the core of SCOOP offers the programmer: a way to spawn off routines asynchronously (all routines invoked on separate objects have this semantics); protection against object-level data races, which by construction cannot occur; a way to explicitly express conditions for condition synchronization by preconditions with wait semantics. These are the main reasons for SCOOP's claim to make concurrent programming "easier", as some concurrency mechanisms are invoked implicitly without the need for programmer statements. This comes at the cost of a runtime system taking care of implicit locking, waiting, etc.

### 2.2 Java Threads

In multithreaded Java[1] (in the following, *Java Threads* for short), no further abstraction level is introduced above threads. Hence in the producer/consumer problem, both the producer and the consumer are threads on their own (inheriting from class *Thread*) and share a buffer as in the following code example:

```
Buffer buffer = new Buffer();
Producer producer = new Producer(buffer);
Consumer consumer = new Consumer(buffer);
```

Once the threads are started

```
producer.start();
consumer.start();
```

the behavior defined in the *run()* methods of *producer* and *consumer* will be executed concurrently.

Mutual exclusion can be ensured by wrapping accesses to the buffer within **synchronized** blocks that mention the object that is used as a lock (in this case *buffer*):

---

[1] We consider "traditional" multithreaded Java, without the higher-level features implemented in later versions of its concurrency library.

```java
public void consume() throws InterruptedException {
    int value;
    synchronized (buffer) {
        while (buffer.size() == 0) {
            buffer.wait();
        }
        value = buffer.get();
    }
}
```

Condition synchronization can be provided by injecting suitable calls to *wait()* and *notify()* methods, which can be invoked on any synchronized object. For example in the *consume()* method, *wait()* is called on *buffer* under the condition that the buffer is empty and puts the calling process to sleep. For proper synchronization, the *notify()* method has in turn to be called whenever it is safe to access the buffer, to wake up any threads waiting on the condition:

```java
public void produce() {
    int value = newValue();
    synchronized (buffer) {
        buffer.put(value);
        buffer.notify();
    }
}
```

In summary, the core of Java Threads offers: a way to define concurrent executions within an object-oriented model; no automatic protection against object-level data races, but a monitor-like mechanism based on **synchronized** blocks; monitor-style *wait()* and *notify()* calls to implement condition synchronization. In comparison with SCOOP, the runtime system is less costly as the programmer is given more responsibility to correctly apply the offered concurrency mechanisms.

## 2.3 Related approaches

Besides the two mentioned models, there are a multitude of concurrent languages, which would also merit comparative studies; we can only mention a few closely related approaches.

High-level concurrency has been proposed for JML [17, 1]; the annotation mechanism works at a different level of abstraction than SCOOP, focusing on method-level locking. An extension of Spec# to multi-threaded programs has been developed [10]; the annotation mechanisms in this extension are very strong, in the sense that they provide exclusive access to objects (making it local to a thread), which may reduce concurrency.

The JCSP approach [22] supports a different model of concurrency for Java, based on the process algebra CSP. JCSP also defines a Java API and set of library classes for CSP primitives, and does not make use of annotations. Polyphonic C# [3] is an annotated version of C# that supports synchronous and asynchronous methods. The language is based on a sound theory (the Join calculus), and is now integrated in the C$\omega$ toolset from Microsoft Research. Morales [15] presents the design of a prototype of SCOOP's separate annotation for Java; however, preconditions and general design-by-contract and support for type safety were not considered. JSCOOP [21] is an attempt to transfer concepts and semantics of SCOOP from its original instantiation in Eiffel to Java.

More generally, a modern abstract programming framework for concurrent or parallel programming is Cilk [4]; Cilk works by requiring the programmer to specify the parts of the program that can be executed safely and concurrently; the scheduler then decides how to allocate work to (physical) processors. Cilk is not yet object-oriented, nor does it provide design-by-contract mechanisms, though recent work has examined extending Cilk to C++.

# 3 Hypotheses

Stating the research questions to be answered is an essential part of the design of any empirical analysis. In the case of our comparative study, a suitable abstract hypothesis is given by the frequently used claim of language designers that programming is simplified by the use of a new language:

> It is easier to program using SCOOP than using Java Threads.

Note that, to support intuition, we explain our study template here and in the following with the concrete languages SCOOP and Java Threads, rather than referring to "Language A", "Language B" etc.; the template is still suitable for general concurrent languages.

A broad formulation such as the above leaves open many possibilities for refinement towards concrete hypotheses:

(1) Program comprehension: Programmers comprehend the meaning of program code better.

(2) Program debugging: Programmers find more errors in program code.

(3) Program correctness: Programmers make fewer errors when writing program code.

(4) Program efficiency: Programmers produce more efficient programs.

(5) Program size: Programmers obtain smaller programs.

(6) Programming speed: Programmers complete a program faster.

Here, suggested hypotheses (1) and (2) recognize the fact that *understanding* programs is typically just as critical in everyday development as being able to produce code. All other suggested hypotheses deal instead with *writing* program code: (3)-(5) are concerned with crucial properties of the programs obtained (program correctness, efficiency, and size), while (6) expresses properties about the process of writing the program (programming speed).

In our study template, we focus on the first three suggested hypotheses. The reason for this is that we feel it is important to ensure correctness before dealing with other properties: measuring efficiency, size, or programming speed is meaningless without ensuring that a correct program is obtained in the first place.

The suggested hypotheses need more refinement, as their formulation fails to capture the connection between the languages under comparison. We work with the following refined concrete hypotheses:

**Hypothesis I** Programmers can comprehend an existing program written in SCOOP more accurately compared to an existing program having the same functionality written in Java Threads (program comprehension).

**Hypothesis II** Programmers can find more errors in an existing program written in SCOOP than in an existing program of the same size written in Java Threads (program debugging).

**Hypothesis III** Programmers make fewer programming errors when writing programs in SCOOP than when writing programs having the same functionality in Java Threads (program correctness).

Note that for the comprehension and correctness tasks we focus on programs having the same functionality, while for the debugging task we require them to have only the same size (close correspondence in number of classes, attributes, functions, and overall lines of code). This is because we want to separate the debugging task from the program's semantics in as far as possible, focusing on "syntactic" errors. This is necessary, as it is impossible to ask for the detection of semantic errors without specifying what the program is supposed to do; providing such a specification would however introduce yet another possibility for misunderstanding.

Note that the list of suggested hypotheses focuses on properties that may be objectively measured: except for program comprehension where it is not obvious on how to obtain a measure (our approach is presented below in Section 6.2), measures are the number of errors, execution speed, lines of code, development time. Other, more subjective, properties are also conceivable (e.g. programmer satisfaction), but not considered in this study.

# 4 Overview of the experimental design

In this section we give an overview of the design of our study; subsequent sections will detail the training phase and the test phase that are part of this design. We start by explaining the basic study setup, using the example of SCOOP vs. Java Threads, then discuss participants' backgrounds in our concrete study.

## 4.1 Setup of the study

As we want to analyze how programming abstractions for concurrency affect comprehension, debugging, and correctness of programs, the study requires human subjects. We have run the study in an academic setting, with 67 students of the Software Architecture course [18] at ETH Zurich in Spring semester 2010. All participants were B.Sc. students, 86.2% in their 4th semester, the others in higher semesters.

This population was split randomly into two groups: the *SCOOP group* (30 students) worked during the study with SCOOP and the *Java group* (37 students) worked with Java Threads. To confirm that the split created groups with similar backgrounds we used both self-assessment and a small number of general proficiency test questions, as detailed below in Section 4.2.

The study had two phases, which we run in close succession of each other: a *training phase*, run during a two-hour lecture session, and a *test phase*, run during an exercise session later on the same day. Two challenges for a study design present themselves:

- *Avoiding bias during the training phase.* We kept the influence by teachers to a minimum through the use of self-study material, discussed further in Section 5.

- *Avoiding bias during the evaluation of the test.* For this we developed a number of objective evaluation schemes, discussed further in Section 6.

In the following we give a brief account of the practical procedure of running the study.

### 4.1.1 Training phase

During the training phase, the participants were given self-study material, depending on their membership in the SCOOP or Java group. The participants were encouraged to work through the self-study material in groups of 2-3 people, but were also allowed to do this individually. The time for working on the study material was limited to 90 minutes. Tutors were available to discuss any questions that the participants felt were not adequately answered in the self-study material.

### 4.1.2 Test phase

During the test phase, participants filled in a pen & paper test, depending on their membership in the SCOOP or Java group. They worked individually, with the time for working on the test limited to 120 min (calculated generously). The tutors of the Software Architecture course invigilated the test and collected the participants' answers at the end of the session.

## 4.2 Student backgrounds

To learn about the students' backgrounds and to confirm that the random split created groups with similar backgrounds we used both self-assessment and a small number of general proficiency test questions; this information was collected during the test phase.

### 4.2.1 Self-assessed programming proficiency

We collected information regarding the current study level of the students and any previous training in concurrency. This confirmed that all students were studying for a B.Sc. degree, and had furthermore taken the 2nd semester Parallel Programming course at ETH, thus starting with similar basic knowledge of concurrency. All students were familiar with Java Threads, as this was the language taught in the Parallel Programming course (we discuss this further in Section 7).

Concerning programming experience we asked the participants to rate themselves on a scale of 5 points where 1 represents "novice" and 5 "expert" regarding their experience in: programming in general; concurrent programming; Java; Eiffel; Java Threads; SCOOP. Figure 1 shows the results with means and standard deviations. Both groups rate their general programming knowledge, as well as their experience with concurrency, Java, and Eiffel at around 3 points, with insignificant differences between the groups. This confirms a successful split of the students into the groups from this self-assessed perspective.

Furthermore, the Java group achieved a higher self-assessed mean for knowledge of Java Threads, and analogous for the SCOOP group. The knowledge of SCOOP, which none of the students was familiar with in the first place, ranked significantly lower than the knowledge of Java Threads.

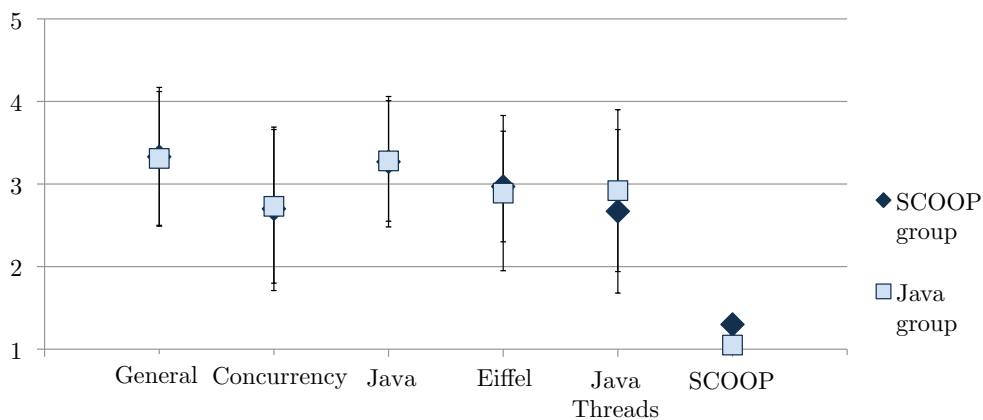

**Figure 1:** Self-assessed programming proficiency

### 4.2.2 General proficiency test

To confirm that the participants have enough knowledge in the base language – Java in the case of Java Threads, and Eiffel in the case of SCOOP – the test included an understanding task: participants were asked for the output of a given program (4 classes, plus an additional wrapper class in Java; approximately 80 lines of code). To assess participants' concurrency knowledge, we also asked multiple choice and text questions on multiprocessing, process states, data races, mutual exclusion, and deadlock.

On both accounts, the students of the two groups achieved very similar results, confirming again the successful split into groups.

## 5  Training phase

When running a comparative study involving novel programming paradigms, study subjects who are proficient in all of these will typically be the exception, making a training phase mandatory. The training process can however also introduce bias, for example if the teaching style of two teachers differs. Requiring the presence of teachers for the study makes it also harder to re-run it, e.g. at other institutions, as a teacher trained in the subject has to be found.

To avoid these problems, we focused on the use of self-study material. Bias could also be introduced when writing this material, but the quality of the material can be judged externally, adding to the transparency of the study, and re-running the study becomes simple.

### 5.1  Self-study material

| Java Threads | SCOOP |
|---|---|
| §1 Concurrent execution | §1 Concurrent execution |
| – Multiprocessing and multitasking | – Multiprocessing and multitasking |
| – Operating system processes | – Operating system processes |
| §2 Threads | §2 Processors |
| – The notion of a thread | – The notion of a processor |
| – Creating threads | – Synch. & asynch. feature calls |
|  | – Separate entities |
| – Joining threads | – Wait by necessity |
| §3 Mutual exclusion | §3 Mutual exclusion |
| – Race conditions | – Race conditions |
| – Synchronized methods | – The separate argument rule |
| §4 Condition synchronization | §4 Condition synchronization |
| – The producer/consumer problem | – The producer/consumer problem |
| – The methods *wait*() and *notify*() | – Wait conditions |
| §5 Deadlock | §5 Deadlock |
| Answers to the exercises | Answers to the exercises |

**Figure 2:** Structure of the self-study material

A course on concurrency can easily take a whole semester. The self-study material we were using and are proposing as a template can be worked through in 90 minutes and thus appears unduly short. However, the material has to be judged in conjunction with the questions of the test and our results in Section 6 show that participants can actually acquire solid basic skills in the limited time frame. A pre-study with six participants, which allowed us to gain various helpful feedback on the study material, confirmed also that the study material can be worked through in 90 minutes.

For teaching the basics of a concurrent language, we suggest the basic structure shown in Figure 2, side-by-side for Java Threads and SCOOP. The only prerequisite for working with these documents is a solid knowledge of the (sequential) base language of the chosen approach, i.e. Java and Eiffel. It is apparent that the documents closely mirror each other, although they describe two different approaches:

§1 This section is identical in both documents, introducing basic notions of concurrent execution in the context of operating systems.

§2 This section concerns the creation of concurrent programs. Here the central notion for Java Threads is that of a thread, for SCOOP it is that of a processor (compare Sections 2.1 and 2.2). At the end of the second section, participants should be able to introduce concurrency into a program, but not yet synchronization.

§3 This section introduces the concept of mutual exclusion. Race conditions and their avoidance using **synchronized** blocks in Java and **separate** arguments in routines in SCOOP are presented.

§4 This section introduces the concept of condition synchronization. The need is explained with the producers/consumers example, and the solutions in Java, i.e. *wait*() and *notify*(), and SCOOP, i.e. execution of preconditions with wait semantics, is explained.

§5 This section introduces the concept of a deadlock.

Furthermore, in every section of the self-study material, there is an equal number of exercises to check understanding of the material; solutions are given at the end of the document.

The SCOOP document had 20 pages including exercises and their solutions, the Java Threads document 18 pages. The self-study material is found in Appendix A (SCOOP) and C (Java Threads).

### 5.2  Students' feedback

To learn about the quality of the training material, we also asked for feedback on the self-study material participants had worked through; this information was collected during the test phase.

Figure 3 gives an overview of the answers to our questions on this topic, rated on a Likert scale of 5 points (where 1 corresponds to "strongly disagree" and 5 to "strongly agree"). Most of the students felt that the material was easy to follow and provided both enough examples and exercises, with insignificant differences between the groups. Both groups also felt that 90 minutes were enough time to work through the material, where the Java group felt significantly better about this point; this might be explained by the fact that the Java group knew some of the material from before. Overall most students agreed, but not strongly, that self-study sessions are a good alternative to traditional lectures.

The overall very positive feedback to the self-study material was confirmed by a number of text comments, and by the tutors invigilating the sessions, who reported that students explicitly expressed that they liked the format of the session.

## 6  Test phase and study results

In this section we present the design of the test and our test evaluation scheme, and report on the results of the concrete study concerning Java Threads vs. SCOOP. We first give further information about student participation, then describe Tasks I to III with their individual evaluation schemes and results, and conclude with a brief interpretation of results. The test material is available in Appendix B (SCOOP) and D (Java Threads).

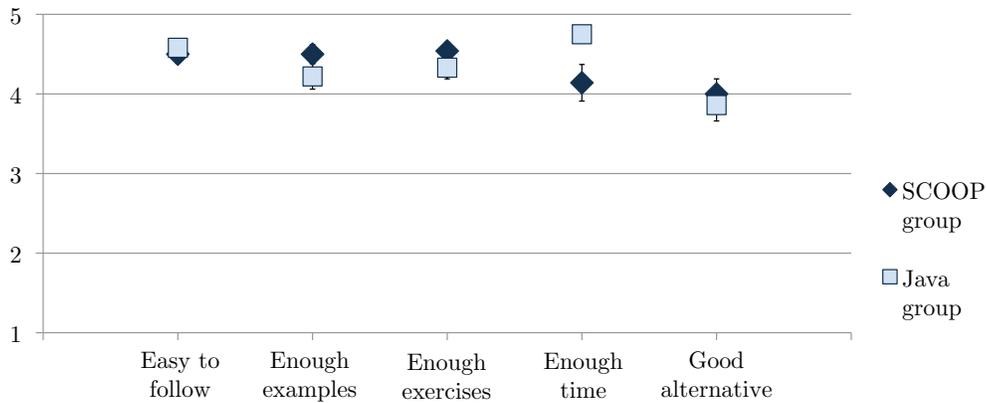

**Figure 3:** Feedback on the self-study material

## 6.1 Student participation and best effort

The participation of the students in the test was high at 84.8% out of 79 students registered in the course. No special incentives such as a prize were given, and the students were told beforehand that their performance in the test cannot affect their grades. Instead, the students were told a week in advance that the lecture and the exercise session on the day of the study would be devoted to the study of two concurrent programming techniques.

In general students filled in all parts of the test diligently, only a minority of 2-5 students per task did not show their best effort by missing very rudimentary answers to Tasks I to III which we filtered out. Concretely, for a solution to be considered as valid

- more than 3 characters (out of 10) per sequence had to be written for Task I;
- at least 1 error (out of 6) had to be found for Task II;
- more than 10 lines of code (out of around 80 for a complete solution) had to be written for Task III.

## 6.2 Task I: Program comprehension

Task I was developed to measure to what degree participants understand the semantics of a program written in a specific paradigm, and thus to test Hypothesis I. Rather than having the semantics described in words, which would make answers ambiguous and their evaluation subjective, we let participants predict samples of a program's output. This task is interesting for concurrent programs, as the scheduling provides nondeterministic variance in the output.

The concrete programs in Java Threads and SCOOP (5 classes, plus an additional wrapper class in Java; ca. 80 lines of code) were printing strings of characters of length 10, with 7 different characters available. In total, the programs' possible outputs contained 28 such sequences, but the participants were neither aware of this number nor the length of the strings. The test asked the participants to write down three of the strings that might be printed by the program.

### 6.2.1 Evaluation

To evaluate the results of Task I, we aimed to find an objective and automatic measure for the correctness of an answer sequence. The obvious measure – stating whether a sequence is correct or not – appeared too coarse-grained. For example, some students forgot to insert a trailing character which was printed after the concurrent computation had finished. Such solutions, although they might show an understanding of concurrent execution as expressed by the language, could have only be marked "incorrect".

We therefore considered the Levenshtein distance [12] as a finer-grained measure, a common metric for measuring the difference between two sequences. In our case, we had to compare not two specific sequences, but a single sequence $s$ with a set $C$ of correct sequences. The algorithm computed the Levenshtein distance $dist$ between $s$ and every element $c \in C$, and took the minimum of the distances:

$$L_{min}(s) = min \ \{dist(s,c) : c \in C\}$$

This corresponds to selecting for $s$ the Levenshtein distance to one of the *closest* correct sequences. As the participants were asked for three such sequences, we took the mean of all three minimal Levenshtein distances to assign a measure to a participant's performance on Task I:

$$\frac{1}{3} \cdot \sum_{i=1,2,3} L_{min}(s_i)$$

**Example 6.1** To illustrate our evaluation algorithm, consider the following example:

| Given sequence | A closest correct sequence | $dist$ |
|---|---|---|
| ATSFTSFPML | ATSFTSFPML | 0 |
| ATSFMTSFPL | ATSFPTSFML | 2 |
| APTSFTSFM | APTSFTSFML | 1 |

In this case we obtain $\frac{1}{3} \cdot (0 + 2 + 1) = 1$. □

### 6.2.2 Results

The results for Task I are displayed in Figure 4 with means and standard deviations.

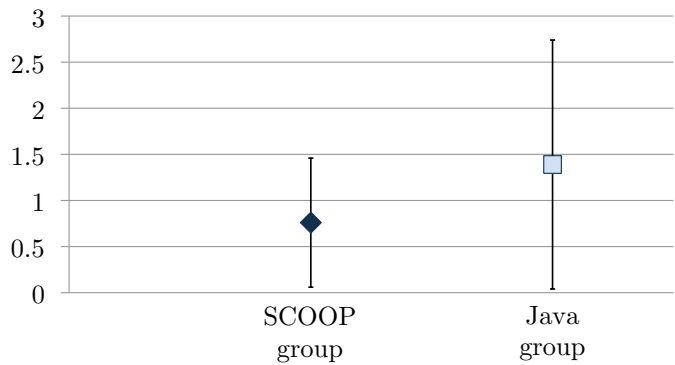

**Figure 4:** Results Task I

A two-tailed independent samples t-test gives that the means can be assumed to be different at a confidence level of 95% (exact significance level 3.3%). This implies that the SCOOP group with the lower mean performed *better* at Task I than the Java group.

## 6.3 Task II: Program debugging

To analyze program debugging proficiency, we provided programs (3 classes, ca. 70 lines of code) that were seeded with 6 bugs. All bugs were of a syntactic nature, meaning that it was not necessary to know the specification of the program to solve the exercise. For Java Threads the bugs included the following types

- Calling *notify()* on a non-synchronized object
- Creating a **synchronized** block without a synchronization object
- Failing to catch an *InterruptedException* for *wait()*

and for SCOOP they included:

- Assigning a **separate** object to a non-separate variable
- Passing a **separate** object as non-separate argument
- Failing to control a **separate** object

Participants were asked for the line of an error, and a short explanation why it is an error.

### 6.3.1 Evaluation

The evaluation assigned every participant points, according to the following scheme:

- 1 point was assigned for pointing out correctly the line where an error was hidden;
- 1 additional point was assigned for describing correctly the reason why it is an error.

The rationale for splitting up the points in this way was that participants may recognize that there is something wrong in a particular line (in this case they would get 1 point), but might or might not know the exact reason that would allow them to fix the error; depending on whether they could actually debug the error, they would get another point.

### 6.3.2 Results

The results for Task II are displayed in Figure 5. A two-tailed independent samples t-test showed a significant difference between the results of the Java and the SCOOP group at a confidence level of 95% (exact significance level 4.2%). This implies that the SCOOP group with the higher mean performed *better* at Task II than the Java group.

## 6.4 Task III: Program correctness

To analyze program correctness, the third task asked participants to implement a program where an object with two integer fields $x$ and $y$ is shared between two threads. One thread continuously tries to set both fields to 0 if they are both 1, the other thread tries the converse. As a pen & paper exercise, the usual compile-time checks that are able to find many of these errors were not available.

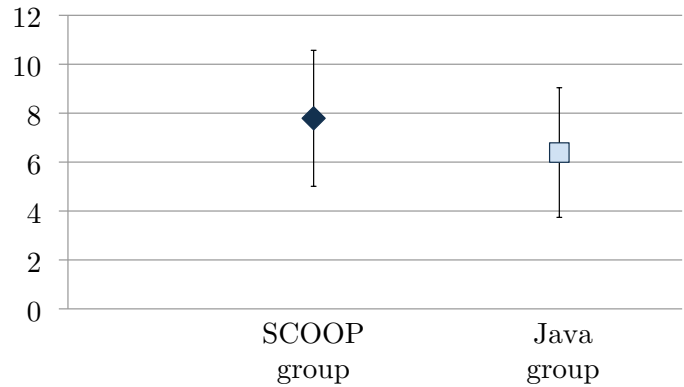

**Figure 5:** Results Task II

### 6.4.1 Evaluation

Even in everyday teaching routine, the grading of a programming exercise can be challenging, and is often not free of subjective influences by the corrector. To avoid such influences in the evaluation of Task III, we used a deductive scheme in which every answer to be graded starts out with 10 points, and points are deducted according to the number and gravity of the errors it contains.

To make this type of grading possible, the grading process was split into several phases:

1. In a first pass of all answers to Task III, attention was paid to the *error types* participants made.

2. The error types were assigned a gravity, which would lead to the deduction of 1 to 3 points.

3. In a second pass of all answers, points were assigned to each answer, depending on the types of errors present in the answer and their gravity.

The gravity of the error was decided as follows:

**Ordinary error** An error that can also occur in a sequential context (1 point deduction).

**Concurrency error** An error that can only arise in a concurrent setting, but which is lightweight as it still allows for concurrent execution (2 points deduction).

**Grave concurrency error** An error that can only arise in a concurrent setting, but is grave as it prevents the program from being concurrent (3 points deduction).

Typos and abbreviations of keywords or other very minor mistakes did not lead to a deduction of points.

### 6.4.2 Error types

The limited size of the programming task led to few error types overall: 7 for Java Threads and 6 for SCOOP. Figures 6 and 7 show the error types with their frequency for Java Threads and SCOOP. Error types with dark/medium/light shaded frequency bars were marked grave concurrency/concurrency/ordinary errors, respectively.

In Java Threads, we considered it a grave error if a proper setup of threads or the starting of threads was missing, hence obtaining a functionless or non-concurrent program. In SCOOP, a direct counterpart to this error was the omission to declare the worker objects *separate*, also leading to a non-concurrent program. 8.3% of Java participants made this error, and 10.7% of SCOOP participants.

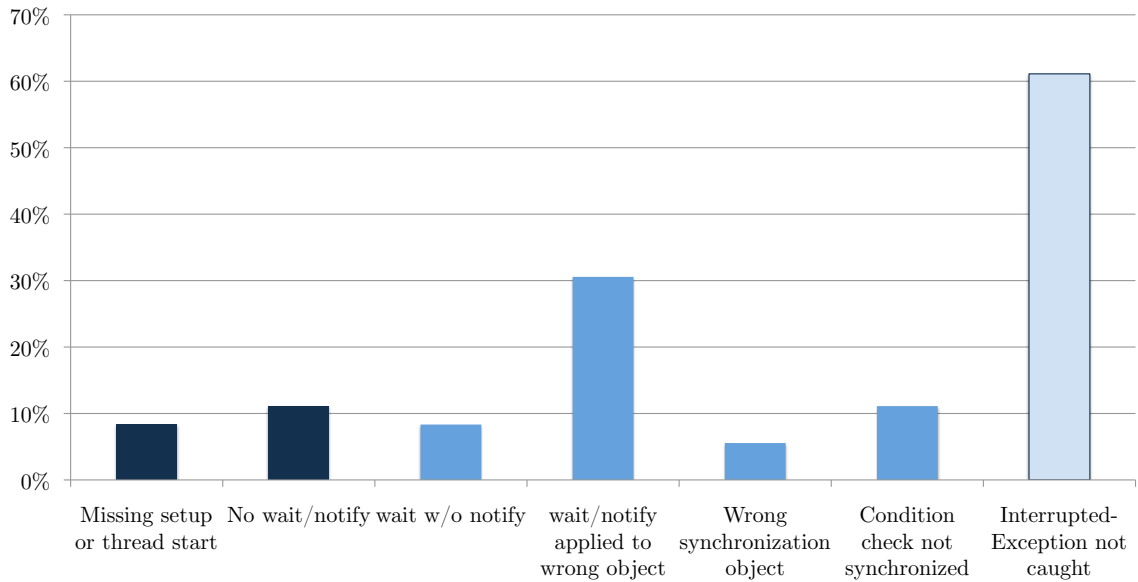

**Figure 6:** Error types for Java Threads

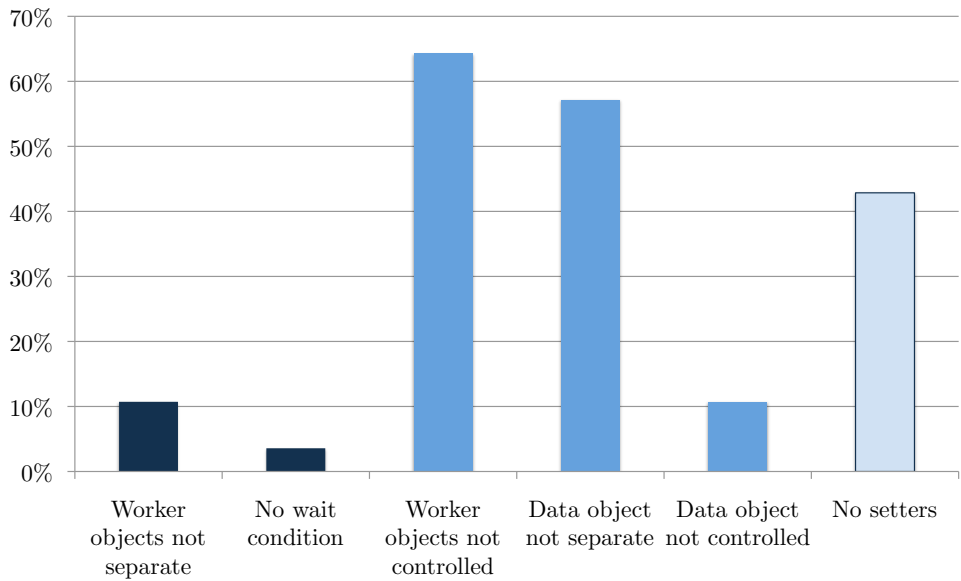

**Figure 7:** Error types for SCOOP

Another grave error was marked for Java Threads if the program did not contain any *wait*() or *notify*() calls, hence providing no condition synchronization. The corresponding error in SCOOP was the absence of wait conditions. Only 3.5% of the SCOOP group made this error, while 11.1% of Java participants did so, an indication that a tighter integration of synchronizing conditions into the programming language might have advantages.

For non-grave concurrency errors and ordinary errors the comparison is no longer that straightforward. A majority of SCOOP participants did not control worker objects and did not declare the data object as *separate*. These are typical novice errors, and would be caught by compile-time checks. Also a large number of SCOOP participants did not use setter routines as needed in Eiffel, a typical ordinary error.

For Java Threads, we see an extreme peak only for not throwing an *InterruptedException* on calling *wait()*, which was classified as an ordinary error and would be caught by compile-time checks. Other concurrency errors involved the use of *wait()* or *notify()*, for example forgetting a corresponding *notify()* or applying it to a wrong object. Note that these errors cannot be caught during compile-time.

### 6.4.3 Results

The results for Task III are displayed in Figure 8. A two-tailed independent samples t-test does not show a significant difference between the two means (exact significance level 32.6%).

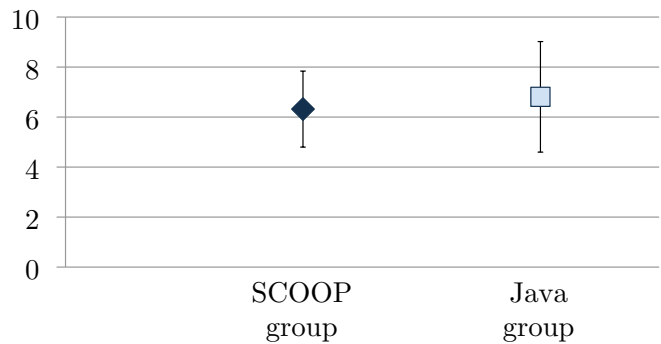

**Figure 8:** Results Task III

## 6.5 Interpretation of the results

The data confirms Hypotheses I and II in favor of SCOOP, leading to the conclusion that SCOOP indeed helps to comprehend and debug concurrent programs. Hypothesis III concerning program correctness could neither be confirmed nor refuted: the SCOOP group did approximately as well as the Java group. Given the small amount of training in the new paradigm, these results are surprising, and promising for the SCOOP model.

The question remains why SCOOP fails to help in program construction. A direct way of interpretation would be to conclude that SCOOP's strengths only affect the tasks of understanding a given program and debugging it. It does not improve constructing correct programs.

However, the first two tasks are at the Comprehension Level of Bloom's taxonomy of learning objectives [2] – level two out of a total of six levels, where a lower level means less cognitively challenging. Comprehension tasks mostly check whether students have grasped how the taught concepts work, an important prerequisite for applying them to new situations. Program construction is at a higher level; depending on the difficulty of presented tasks and previously studied examples, it could be on one of the level three to five of Bloom's taxonomy. It is possible that the training time allotted for this study was too short to enable students transfer the abstractions to the new problem presented in the test. To find out whether this was the case and SCOOP, in comparison to Java Threads, also benefits program construction, a re-run of the study with a more extensive training phase would be necessary.

# 7 Threats to validity

The fact that all students of our study had previous knowledge of Java Threads, but none of SCOOP, can be expected to skew the results to benefit Java Threads. We were aware of this situation already in the planning phase of the study, and decided to run it with this group of participants nonetheless. A similar situation also frequently arises in practice: developers versed in a certain programming paradigm consider learning a new one. The study results show that even under these circumstances, the new paradigm might prove superior to the well-known one (Tasks I and II).

Another threat to internal validity is the experimenter bias, where the experimenter inadvertently affects the outcome of the experiment. A double-blind study was not an option in our case, as at least some of the results had to be analyzed by humans, at this time revealing the membership to a group in the experiment. Using automatic techniques for Task I, clearly defined errors with line numbers in Task II, and developing the deductive scheme for Task III should however limit this bias to a minimum.

A further threat to internal validity is that results might have been influenced by the usability of the base programming languages themselves, Java and Eiffel. In self-assessment participants attributed themselves however sufficient proficiency in both languages and this was confirmed by a short test (see Section 4.2); these influences might thus be negligible.

As a threat to external validity, we used only students as study subjects and it is unclear how the study results generalize to other participant groups and situations. In particular, the use of development environments might greatly affect the learning experience and the potential of producing correct programs. We suggest to run further studies in the future (see Section 9.2) to explore these situations, but deem our study a "cleanroom approach" to analyzing the effects of language abstractions.

As a threat to construct validity, it is difficult to justify objectively that tasks were "fair" in the sense that they did not favor one approach over the other. However, Java Threads and SCOOP are languages that are suitable for ordinary concurrency tasks, and such tasks featured in the test. This situation would be more difficult for languages that aim for a specific application domain.

# 8 Related work

According to Wilson et al. [23], the evaluation of parallel programming systems should encompass three main categories of assessment factors: a system's run-time performance, its applicability to various problems, and its usability (ease of learning and probability of programming errors).

The assessment of the factors described in the first two categories are directly related to metrics that can be collected through, for example, running benchmark test suites. But, as shown for the domain of modeling languages by Kamandi et al. [11], metrics cannot predict the outcomes of controlled experiments with human subjects for the assessment factors of the third category "usability".

The need for controlled empirical experiments for concurrent programming has already been recognized 15 years ago [19]. Nevertheless, only few such experiments have been carried out so far. Those that have been carried out focus on time that it takes the study participants to complete a given programming assignment.

Szafron and Schaefer [19] conducted an experiment with 15 students of a concurrent programming graduate course. They taught two parallel programming systems (one high-level system and a message-passing library system) each for 50 minutes to the entire class; students then had two weeks to solve a programming assignment in a randomly assigned system. The evaluation compared the time students worked on the assignment, number of lines, and run-time speed amongst other measures. Their results suggest that the high-level system is more usable the message passing library, although students spent more time on the task with the high-level system.

The group around Hochstein, Basili, and Carver conducted multi-institutional experiments [9, 8, 7] in the area of high performance computing using parallel programming assignments and students as subjects. In all these experiments, time to completion is the main measure taken. The results of these studies indicate that the message passing approach to parallel programming takes more total effort than the shared memory approach.

Luff [13] compares the programmer effort using traditional lock-based approaches to the Actor model, and transactional memory systems. He uses time taken to complete a task and lines of code as objective measures and a questionnaire capturing subjective preferences. The data exhibits no significant differences based on the objective measures, but the subjective measures show a significant preference of the transactional memory approach over the standard threading approach.

All of the above experiments target programmer productivity as their main focus. To measure this, the studies need to provide substantial programs and a long time range for completing them as a basis of work. By doing so, some of the control over the experimental setup is lost. Our study has a more modest goal: it tries to compare two approaches with respect to their ease of learning them and understanding and writing small programs correctly after a very short time of instruction. By narrowing the focus in such a way, we place the ability of controlling the experiment over being able to generalize the results to arbitrary situations and levels of proficiency. Given that this experiment is only a first step in a series, it seems justified to do so.

Other studies [5, 6] consider more generally the comparison of programming paradigms, without a focus on concurrency. The study of Carey and Shepherd [5] focused on learning new paradigms and how students are affected by their past experience. Harrison et al. [6] compared functional programming to object oriented programming. The problem with their experimental approach is their use of only a single developer to implement program with the same functionalities in both C++ and SML. They did not detect significant differences in the number of errors found, but they showed that SML programs had more use of library routines and took longer to test.

# 9 Conclusion

## 9.1 Discussion

The use of programming abstractions since the 1960s has enabled the tremendous growth of computing applications witnessed today. New challenges such as multicore programming await the developers and the languages community, but the multitude of proposals makes it hard for a new language to leave a mark. Empirical studies are urgently needed to be able to judge which approaches are promising. Since abstractions are invented for the sake of the human developer, and to finally improve the quality of written code, such studies have to involve human subjects.

Despite the need for such studies, they have been run only infrequently in recent years. One reason for this might be that there is too much focus on established languages. Hence newly proposed languages are not put to the test as they should, ultimately hampering the progress of language research. For this reason we have proposed a template for a study, which can expressly be used with novel paradigms. While established study templates are a matter of course in other sciences, they are not common (yet) in empirical software engineering. We also feel that the research community should focus their attention on developing templates, as they will improve research results in the long term and provide a higher degree of comparability among studies.

The key to making our study template successful was the reliance on self-study material in conjunction with a test, and an evaluation scheme that exposes subjective decisions of the corrector. While 90 minutes for studying a new language is brief, we were actually impressed how much the participants learned, some of which handed in flawless pen & paper programs.

## 9.2 Future work

Clearly, our template should be applied to more languages in the future; there is an abundance of them, as discussed in Section 2. Also, the set of study subjects can be varied in future studies. In an academic setting, we would ideally like to re-run the Java/SCOOP study with students who have no prior concurrency experience. Also, the study template should be used at other institutions, and in the end grow out of the academic setting and involve developers.

The template could also be developed further. For example, it would be possible to concentrate more strongly on one aspect, e.g. program correctness, and to pose more tasks to test a single hypothesis. The evaluation in Section 6.4 shows that participants might have improved their results greatly if they have had access to a compiler; running the test not as a pen & paper exercise but with computer support would thus be yet another option.

# Acknowledgments

We would like to thank S. Easterbrook and M. Chechik for providing valuable comments and suggestions on this work. We thank A. Nikonov, A. Rusakov, Y. Pei, L. Silva, M. Trosi, and J. Tschannen, who provided helpful feedback on the study material as part of a pre-study; B. Morandi and S. West, who were tutors during the self-study session; M. Nordio, S. van Staden, J. Tschannen, and Y. Wei, who were tutors during the test session.

This work is part of the SCOOP project at ETH Zurich, which has benefited from grants from the Hasler Foundation, the Swiss National Foundation, Microsoft (Multicore award), ETH (ETHIIRA). F. Torshizi has been supported by a PGS grant from NSERC.

# A  SCOOP: Self-study material

Almost all computer systems on the market today have more than one CPU, typically in the form of a multi-core processor. The benefits of such systems are evident: the CPUs can share the workload amongst themselves by working on different instructions in parallel, making the overall system faster. This work sharing is unproblematic if the concurrently executing instructions are completely independent of each other. However, sometimes they need to access the same region of memory or other computing resources, which can lead to so-called *race conditions* where the result of a computation depends on the order of nondeterministic system events. Therefore concurrent processes have to be properly *synchronized*, i.e. programmed to wait for each other whenever necessary, and this calls for specialized programming techniques.

Today you will learn about the background and techniques of *concurrent programming*. In particular, you will get to know an object-oriented programming model for concurrency called *SCOOP* (Simple Concurrent Object-Oriented Programming). At the end of this lesson, you will be able to

- explain the basics of concurrent execution of processes in modern operating systems, in particular multiprocessing and multitasking,

- understand some of the most important problems related to concurrent programming, in particular race conditions and deadlocks,

- distinguish between different types of process synchronization, in particular mutual exclusion and condition synchronization,

- understand how these types of synchronization are realized in the SCOOP programming model,

- program simple concurrent programs using SCOOP.

The lesson consists entirely of self-study material, which you should work through in the usual two lecture hours. You should have a study partner with whom you can discuss what you have learned. At the end of each study section there will be exercises that help you test your knowledge; solutions to the exercises can be found on the last pages of the document.

## A.1  Concurrent execution

This section introduces the notion of concurrency in the context of operating systems. This is also where the idea of concurrent computation has become relevant first, and as we all have to deal with operating systems on a daily basis, it also provides a good intuition for the problem. You may know some of this content already from an operating systems class, in which case you should see this as a review and check that you are familiar again with all the relevant terminology.

### A.1.1  Multiprocessing and multitasking

Up until a few years ago, building computers with multiple CPUs (Central Processing Units) was almost exclusively done for high-end systems or supercomputers. Nowadays, most end-user computers have more than one CPU in the form of a multi-core processor (for simplicity, we use the term CPU also to denote a processor core). In Figure 16 you see a system with two CPUs, each of which handles one process.

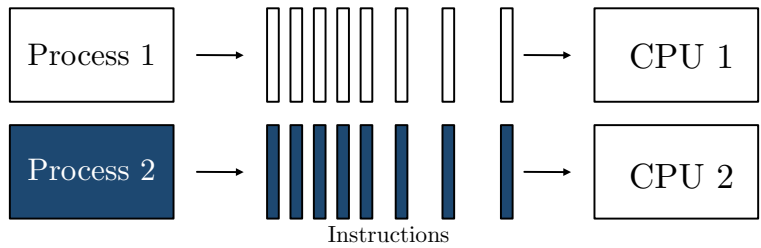

**Figure 9:** Multiprocessing: instructions are executed in parallel

The situation where more than one CPU is used in a single system is known as *multiprocessing*. The processes are said to execute *in parallel* as they are running at the same time.

However, also if you have a computer with a single CPU, you may still have the impression that programs run "in parallel". This is because the operating system implements *multitasking*, i.e. makes a single CPU appear to work at different tasks at once by switching quickly between them. In this case we say that the execution of processes is *interleaved* as only one process is running at a time. This situation is depicted in Figure 17. Of course, multitasking is also done on multiprocessing systems, where it makes sense as soon as the number of processes is larger than the number of available CPUs.

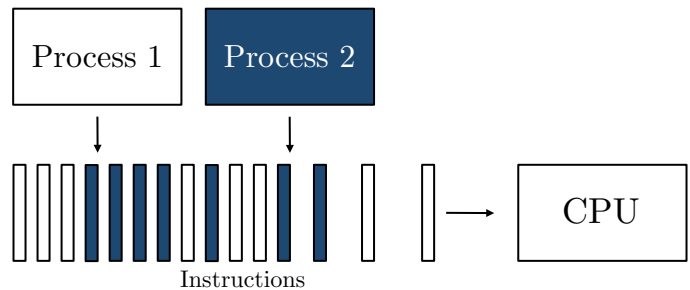

**Figure 10:** Multitasking: instructions are interleaved

Both multiprocessing and multitasking are examples of *concurrent execution*. In general, we say that the execution of processes is *concurrent* if they execute either truly in parallel or interleaved. To be able to reason about concurrent executions, one often takes the assumption that any parallel execution on real systems can be represented as an interleaved execution at a fine enough level of granularity, e.g. at the machine level. It will thus be helpful for you to picture any concurrent execution as the set of all its potential interleavings. In doing so, you will be able to detect any inconsistencies between different executions. We will come back to this point in Section C.3.1.

In the following section we will see how operating systems handle multitasking, and thus make things a bit more concrete.

### A.1.2   Operating system processes

Let's have a closer look at processes, a term which we have used informally before. You will probably be aware of the following terminology: a (sequential) *program* is merely a set of instructions; a *process* is an instance of a program that is being executed. The exact structure of a process may change from one operating system to the other; for our discussion it suffices to assume the following components:

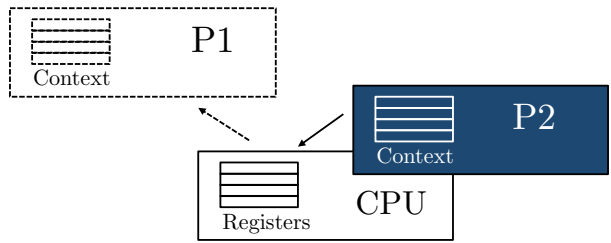

**Figure 11:** Context switch: process P1 is removed from the CPU and P2 is assigned to it

- *Process identifier*: the unique ID of a process.
- *Process state*: the current activity of a process.
- *Process context*: the program counter and the values of the CPU registers.
- *Memory*: program text, global data, stack, and heap.

As discussed in Section C.1.1, multiple processes can execute at the same time in modern operating systems. If the number of processes is greater than the number of available CPUs, processes need to be scheduled for execution on the CPUs. The operating system uses a special program called the *scheduler* that controls which processes are *running* on a CPU and which are *ready*, i.e. waiting until a CPU can be assigned to them. In general, a process can be in one of the following three states while it is in memory:

- *running*: the process's instructions are executed on a processor.
- *ready*: the process is ready to be executed, but is not currently assigned to a processor.
- *blocked*: the process is currently waiting for an event.

The swapping of process executions on a CPU by the scheduler is called a *context switch*. Assume a process P1 is in the state *running* and should be swapped with a process P2 which is currently *ready*, and consider Figure 18. The scheduler sets the state of P1 to *ready* and saves its context in memory. By doing so, the scheduler will be able to wake up the process at a later time, such that it can continue executing at the exact same point it had stopped. The scheduler can then use the context of P2 to set the CPU registers to the correct values for P2 to resume its execution. Finally, the scheduler sets P2's process state to *running*, thus completing the context switch.

From the state *running* a process can also get into the state *blocked*; this means that it is currently not ready to execute but waiting for some system event, e.g. for the completion of some prerequisite task by another process. When a process is *blocked* it cannot be selected by the scheduler for execution on a CPU. This can only happen after the required event triggers the state of the blocked process to be set to *ready* again.

**Exercise A.1** Explain the difference between parallel execution, interleaved execution, and concurrent execution.

**Exercise A.2** What is a context switch? Why is it needed?

**Exercise A.3** Explain the different states a process can be in at any particular time.

## A.2 Processors

Concurrency seems to be a great idea for running different sequential programs at the same time: using multitasking, all programs appear to run in parallel even on a system with a single CPU, making it more convenient for the user to switch between programs and have long-running tasks complete "in the background"; in the case of a multiprocessing system, the computing power of the additional CPUs speeds up the system overall.

Given these conveniences, it also seems to be a good idea to use concurrency not only for executing different sequential programs, but also within a single program. For example, if a program implements a certain time-intensive algorithm, we would hope that the program runs faster on a multiprocessing system if we can somehow *parallelize* it internally. A program which gives rise to multiple concurrent executions at runtime is called a *concurrent program*.

### A.2.1 The notion of a processor

Imagine the following routine *compute* which implements a computation composed of two tasks:

>  *compute*
>     **do**
>        *t1.do_task1*
>        *t2.do_task2*
>     **end**

Assume further that it takes $m$ time units to complete the call *do_task1* on the object attached to entity *t1* and $n$ time units to complete *do_task2* on the object attached to entity *t2*. If *compute* is executed sequentially, we thus have to wait $m$ time units after the call *t1.do_task1* before we can start on *t2.do_task2*, and the overall computation will take $m + n$ time units, as shown in Figure 19.

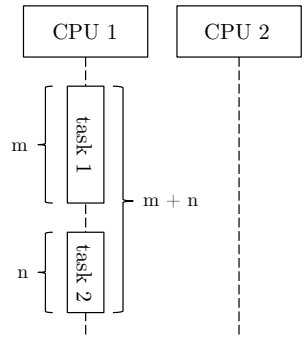

**Figure 12:** Sequential execution: the overall computation takes $m + n$ time units

If we have two CPUs, this seems rather a waste of time. What we would like to do instead is to execute *do_task1* on the object attached to entity *t1* by one of the CPUs and *do_task2* on the object attached to entity *t2* by the other CPU, such that the overall computation takes only $max(m, n)$ time units, as shown in Figure 20.

In order to be able to associate computation with different execution units, we introduce the abstract notion of a processor. An (abstract) *processor* can best be understood as a kind of virtual CPU: an entity supporting sequential execution of instructions on one or several objects. Each processor has a *request queue* which holds the instructions that it has to execute, and works them off one by one.

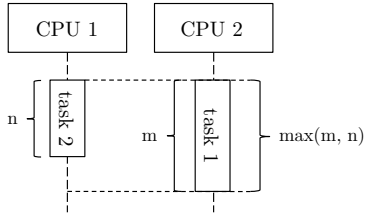

**Figure 13:** Parallel execution: the overall computation takes $max(m, n)$ time units

In contrast to physical CPUs, the number of processors is not bounded. We imagine that processors can be assigned to physical CPUs via multitasking, just as operating systems processes are. In the following we will use the term processor only in this abstract sense, and use the term CPU to denote a physical CPU.

The fundamental idea of abstract processors in SCOOP is their relationship to objects: each object is assigned to exactly one processor, called the *handler* of the object. On the other hand, a processor can handle multiple objects.

If a new object is created, the runtime system decides which handler it is assigned to or whether a new processor is created for it, and this assignment remains fixed over the course of the computation. The assignment is guided by an extension of the type system, as we will see later. Assume for now that *t1* is handled by a processor *p*, and *t2* and *t3* are handled by a processor *q*. We can depict this with the diagram shown in Figure 14.

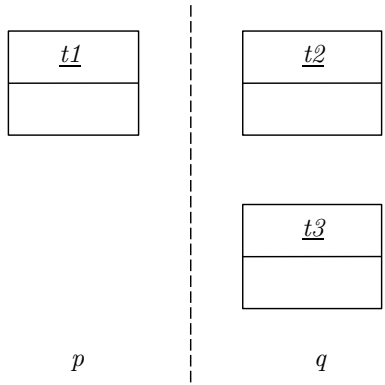

**Figure 14:** Processor regions: *t1* is handled by processor *p*, and *t2* and *t3* are handled by a processor *q*

We frequently use such diagrams as they give us an idea of the associations of processors and objects. Each region tagged by a processor name contains the objects this processor is handling; processor regions are separated by a dashed line.

### A.2.2 Synchronous and asynchronous feature calls

What does the *handling* of an object imply? It means that all operations on the given object are executed by its handling processor; there is no sharing of objects between processors. For example, assume that the following feature calls

*t1.do_task1*
*t2.do_task2*
*t3.f*

are encountered by the current processor $q$, and that the processor association are as in Figure 14. Then $q$ doesn't execute the call *t1.do_task1* by itself, but asks $p$ to do it by appending the call to $p$'s request queue. The benefit is that processor $q$ doesn't have to wait for the call *t1.do_task1* to complete: in contrast to the sequential case, $q$ can just continue with other calls, e.g. *t2.do_task2*, which it is handling by itself. Hence the two tasks can be executed concurrently. Lastly, the call *t3.f* is handled once again by processor $q$, therefore it is only started after the task *t2.do_task2* has been completed.

A feature call on an object which is handled by a different processor than the current one is called an *asynchronous* feature call or a *separate call* (e.g., *do_task1*). In this case the current processor can proceed after issuing the call to the other processor, and doesn't have to wait for the call to return. In contrast, a feature call on an object handled by the current processor is called a *synchronous* feature call or a *non-separate call* (e.g. *do_task2*). This is the situation well-known from ordinary sequential programming, and the current processor has to wait for the call to return before continuing.

### A.2.3  Separate entities

We have left open the question of how the runtime system determines whether a particular object is handled by one processor or another. The answer is that the type system is extended to guide the runtime system in this decision, thus giving the programmer control over whether a call is executed synchronously or asynchronously.

To this end, a new keyword is introduced in the language SCOOP: **separate**. Along with the usual

$x : X$

to denote an entity $x$ that can be attached to objects of type $X$, we can now also write

$x :$ **separate** $X$

to express that at runtime, $x$ may be attached to objects handled by a different processor. We then say that $x$ is of type **separate** $X$, or that it is a *separate entity*.

The value of a separate entity is called a *separate reference*, and an object attached to it is called a *separate object*. To emphasize that a certain reference or object is not separate, we use the term *non-separate*. We also extend our diagrams to include references to objects by drawing arrows to the referenced object. If an arrow crosses the border of a processor's domain, it is a separate reference. The diagram in Figure 15 shows two objects $x$ and $y$ which are handled by different processors, where $x$ contains a separate reference to $y$.

### A.2.4  Wait-by-necessity

We generalize the example of Section A.2.1 by defining the following class *WORKER*:

    **class** *WORKER*
        **feature**
            *output*: **INTEGER**
            *do_task* (*input*: **INTEGER**) **do** ... **end**
    **end**

The idea is that a worker can run *do_task* (*input*) and will store the result in the feature *output*. Let's assume that two workers are defined in the class *MANAGER* as follows:

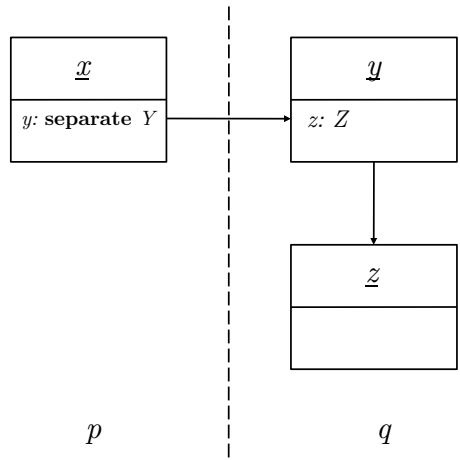

**Figure 15:** Separate reference: *y* references an object on a different processor

```
class MANAGER
    feature
        worker1 : separate WORKER
        worker2 : WORKER

        -- in some routine:
            do
                . . .
                worker1.do_task (input1)
                worker2.do_task (input2)
                result := worker1.output + worker2.output
            end
    end
```

We have learned before that separate calls are spawned off asynchronously, so the current processor doesn't have to wait for them to return. The call *worker1.do_task (input1)* is therefore executed on a different processor; the second call *worker2.do_task (input2)* is synchronous and is executed on the same processor the manager object is running on. Note that the call *worker1.do_task (input1)* is a *command* and thus just transforms the target object *worker1*, without returning a result. But what about the call *worker1.output*? This is a *query* and thus returns a result. As we are interested in the result, we clearly have to wait for the call to return; furthermore, we would also like to wait until previous computations on the object are finished before retrieving the information.

This waiting happens automatically in SCOOP, and the corresponding synchronization principle is called *wait-by-necessity*:

> "If a client has started one or more calls on a certain separate object, and it executes on that object a call to a query, that call will only proceed after all the earlier ones have been completed, and any further client operations will wait for the query to terminate."

This rule ensures that after completion of the call *worker2.do_task (input2)*, the processor will also wait for the asynchronous completion of call *worker1.do_task (input1)* before combining the results. This mechanism is completely automatic, so you as a programmer don't have to worry about this. However, when trying to optimize programs, it is important to know that queries on an object will act as a *barrier*, i.e. a program point where execution waits for all previously spawned calls on that object before proceeding.

**Exercise A.4** How does the execution of an asynchronous feature call differ from a synchronous one? How are asynchronous feature calls expressed in SCOOP?

**Exercise A.5** Consider that the following SCOOP program fragment is executed on a processor $p$:

*worker1.do_task1*
*worker2.do_task2*
*manager.evaluate*
*worker3.do_task3*
*result := worker2.value + worker3.value*
*manager.finish*

The object-processor associations are given as follows: *worker1* and *worker2* are handled by processor $q$, *manager* by processor $p$, and *worker3* by processor $r$. The call *worker1.do_task1* takes 20 time units until it returns, *worker2.do_task2* 30 time units, *manager.evaluate* 40 time units, *worker3.do_task3* 20 time units, *manager.finish* 20 time units; the queries return immediately. What is the minimum time for execution of this program? Draw a sequence diagram to justify your answer.

**Exercise A.6** Consider classes $A$ and $B$

| **class** $A$ | **class** $B$ |
|---|---|
| **feature** | **feature** |
|     $b$: **separate** $B$ |     $a$: **separate** $A$ |
|     $c$: $C$ | |
| |     *set_a* ($a1$: **separate** $A$) **do** $a := a1$ **end** |
|     *set_b* ($b1$: **separate** $B$) **do** $b := b1$ **end** | |
|     *set_c* ($c1$: $C$) **do** $c := c1$ **end** | |
| **end** | **end** |

and assume that the following program fragment is executed

*a.set_b* (*b*)
*a.set_c* (*c*)
*b.set_a* (*a*)

where $a$ and $c$ are handled by processor $p$, and $b$ is handled by processor $q$. Draw a diagram showing the association of objects with processor regions and any separate or non-separate references.

**Exercise A.7** Under what conditions does wait-by-necessity become applicable?

## A.3 Mutual exclusion

Up until now, concurrency seems easy enough to handle. If we want a feature to be evaluated concurrently, we have to declare its corresponding target **separate**. At runtime, this gives rise to an asynchronous feature call, and we are done. However, what happens if different calls interfere with each other, for example access and modify the same objects? We will see that this might change the results of computations in unexpected ways, and we thus have to avoid these situations by using a special type of synchronization called *mutual exclusion*. Luckily, SCOOP has a simple mechanism for ensuring mutual exclusion.

### A.3.1 Race conditions

Consider the following class *COUNTER* which only has a single attribute *value*, and features to set and increment *value*.

```
class COUNTER
    feature
        value : INTEGER

        set_value (a_value: INTEGER)
            do
                value := a_value
            end

        increment
            do
                value := value + 1
            end
end
```

Now assume that an entity $x$ of type **separate** *COUNTER* is created and consider the following code:

$x.set\_value\ (0)$
$x.increment$
$i := x.value$

What is the value of $i$ at the end of this execution? Clearly, if this code was part of a sequential program, the value would be 1. In a concurrent setting where we have two or more processors, the value of $x$ can be read/modified by all processors though that handle objects owning a separate reference to $x$. For example consider the following call executed concurrently by another processor (different from the processor executing the above code):

$x.set\_value\ (2)$

What is the value of $i$ now?

The answer is that, if these are the only feature calls running concurrently and $x$ is attached to the same object in both cases, $i$ could have any of the values 1, 2, or 3. The reason for this is easily explained. Assume that processor $p$ is handling the object associated with $x$. This processor will receive feature calls for evaluation from concurrently executed code parts, and will interleave them. The following interleavings could be taken:

| $x.set\_value\ (2)$ | $x.set\_value\ (0)$ | $x.set\_value\ (0)$ | $x.set\_value\ (0)$ |
| $x.set\_value\ (0)$ | $x.set\_value\ (2)$ | $x.increment$ | $x.increment$ |
| $x.increment$ | $x.increment$ | $x.set\_value\ (2)$ | $i := x.value$ |
| $i := x.value$ | $i := x.value$ | $i := x.value$ | $x.set\_value\ (2)$ |
| $i = 1$ and $x.value = 1$ | $i = 3$ and $x.value = 3$ | $i = 2$ and $x.value = 2$ | $i = 1$ and $x.value = 2$ |

This is not really what we intended. The result of our computation has become arbitrary, and depends on the scheduling that determines a particular interleaving. Remember that we have no control over the scheduling.

The situation that the result of a concurrent execution is dependent on the nondeterministic scheduling is called a *race condition* or a *data race*. Data races are one of the most prominent problems in the domain of concurrent programming, and you can imagine that it gives rise to errors which can be quite hard to detect. For example, when you are running a program such as the above, say, 100 times, it might be that, because of a specific timing of events, you always obtain the values $i = 1$ and $x.value = 1$. But when you run the program for the 101st time, one of the other results arises. This means that such errors can stay hidden for a long time, and might never be detected during testing.

The question is now how to avoid data races. SCOOP has a specific mechanism for this that eliminates these types of errors at compile-time (before you even run the program!), which will be explained in the next section.

### A.3.2 The separate argument rule

To avoid data races we have to *synchronize* different computations such that they don't interfere with each other. Let's think about the main reason for the problem to occur. In the above example, two computations *shared* a resource, namely the object attached to $x$. A part of a program that accesses a shared resource is called a *critical section*. The problem would not have occurred if, at any time, at most one computation would be in its critical section. The form of synchronization ensuring this property is called *mutual exclusion*.

SCOOP has a simple way to ensure mutual exclusion: its runtime system automatically *locks* the processors which handle separate objects passed as arguments of a routine. If a processor is locked, no other computation can use it to evaluate a feature call; the processor becomes private to whoever locked it. Let's make an example to see how that helps us.

Recall the above example, but let's extend it to see the routine the code has been taken from:

>   *compute* (*x*: **separate** *COUNTER*)
>       **do**
>           *x.set_value* (0)
>           *x.increment*
>           *i* := *x.value*
>       **end**

Consider now the call *compute* (*x*) and assume that $x$ is handled by processor $p$. As explained above, since $x$ is a separate argument to the routine, the processor $p$ must be locked. The current processor, which is about to execute the *compute* feature, waits until the underlying runtime system locks processor $p$. As soon as $p$ is locked, the body of the routine can be executed without interference (multiple locks on a processor are not possible), and hence upon completion of the routine we always obtain the result $i = 1$ and $x.value = 1$.

This is so important that SCOOP forces us to make separate entities which we want to access an argument of the enclosing routine. This is formulated as the *separate argument rule*:

> "The target of a separate call must be a formal argument of the routine that contains the separate call."

In other words, all calls on separate objects must be wrapped in a procedure that makes it possible to pass the target as argument. Hence only one of the following two examples is correct:

| $x$ : **separate** $X$<br>*compute*<br>    **do**<br>        *x.f*<br>    **end** | $x$ : **separate** $X$<br>*compute* ($x1$: **separate** $X$)<br>    **do**<br>        *x1.f*<br>    **end** |
|---|---|
| *Incorrect*: Target $x$ is declared separate, but not an argument of the enclosing routine *compute*. | *Correct*: Target $x1$ is separate and therefore has to be an argument of the enclosing routine. In order to execute *x.f*, we use the call *compute* ($x$). |

Note that if an argument of a separate type is passed, the corresponding formal argument must also be of separate type. Thus, in the example above on the right hand side, $x1$ must be declared of type **separate** $X$, since $x$, which we want to pass as an argument is also of type **separate** $X$. This type system restriction avoids that entities declared as non-separate can become attached to separate objects, which would compromise the correctness of the SCOOP model.

An analogous requirement holds also for assignments. For example, if $x1 := x$ and $x$ is of type **separate** $X$ (or might just be attached to an object on a separate processor), then $x1$ must be of type **separate** $X$ too. This can be remembered by "*nonsep* := *sep*" being disallowed, and is also summarized in the following typing rule:

> "If the source of an attachment (assignment or argument passing) is separate, its target must be separate too."

Note that an assignment the other way around ( "*sep* := *nonsep*", i.e. non-separate source, separate target) is however admissible.

**Exercise A.8** Explain the terms data race and mutual exclusion. How does SCOOP ensure mutual exclusion?

**Exercise A.9** Consider the following class *MOTORBIKE* that models a motorbike with engine, wheels, and status display. The class doesn't compile properly. Find all errors and fix them, assuming that the type declarations of all attributes are correct and that the omitted classes *ENGINE*, *DISPLAY*, *WHEEL* have the features mentioned.

    **class** *MOTORBIKE*
        **create**
            *make*
        **feature**
            *engine*: **separate** *ENGINE*
            *display*: *DISPLAY*
            *front_wheel*: **separate** *WHEEL*
            *back_wheel*: **separate** *WHEEL*

            *make*
                **do**
                    **create** *engine*; **create** *display*
                    **create** *front_wheel*; **create** *back_wheel*
                **end**
            *initialize*

```
        do
            engine.initialize
            initialize_wheels
            display.show ("Ready")
        end

    initialize_wheels
        do
            display.show ("Initializing wheels ...")
            front_wheel.initialize
            back_wheel.initialize
        end

    switch_wheels
        local
            wheel: WHEEL
        do
            wheel := front_wheel
            front_wheel := back_wheel
            back_wheel := wheel
        end
end
```

## A.4 Condition synchronization

Protecting access to shared variables is not the only reason why a process has to synchronize with other processes. For example, assume that a process continuously takes data items out of a buffer to process them. Hence, the process should only access the buffer if it holds at least one element; if it finds the buffer empty, it therefore needs to wait until another process puts a data item in. Delaying a process until a certain condition holds (as in this case, until the "buffer is not empty") is called *condition synchronization*. As you will see, SCOOP has an elegant way of expressing condition synchronization by reinterpreting the preconditions of a routine as *wait conditions*.

As an example of a problem that requires processes to use condition synchronization, we describe the so-called *producer-consumer problem*, which corresponds to issues found in many variations on concrete systems. Devices and programs such as keyboards, word processors and the like can be seen as *producers*: they produce data items such as characters or files to print. On the other hand the operating system and printers are the *consumers* of these data items. It has to be ensured that these different entities can communicate with each other appropriately so that for example no data items get lost.

On a more abstract level, we can describe the problem as follows. We consider two types of processes, both of which execute in an infinite loop:

- *Producer*: At each loop iteration, produces a data item for consumption by a consumer.

- *Consumer*: At each loop iteration, consumes a data item produced by a producer.

Producers and consumers communicate via a shared buffer implementing a queue; we assume that the buffer is unbounded, thus we only have to take care not to take out an item from an empty buffer, but are always able to insert new items. Instead of giving the full implementation we just assume to have a generic class $BUFFER[T]$ to implement an unbounded queue:

    *buffer*: **separate** $BUFFER[\textbf{INTEGER}]$

Producers append data items to the back of the queue using a routine *put*(*item*: **INTEGER**), and consumers remove data items from the front using *get*: **INTEGER**; the number of items in a queue is determined by the feature *count*: **INTEGER**.

As part of the consumer behavior, we might for example want to implement the following routine for consuming data items from the buffer:

    *consume* (*a_buffer*: **separate** $BUFFER[\textbf{INTEGER}]$)
        **require**
            **not** (*a_buffer.count* == 0)
        **local**
            *value*: **INTEGER**
        **do**
            *value* := *a_buffer.get*
        **end**

Note that we have used a precondition to ensure that if we attempt to get a value from the buffer, it is not currently empty. However, what should happen if the buffer is indeed found empty? In a sequential setting, we would just throw an exception. However, this is not justified in the presence of concurrency: eventually a producer will put a value into the buffer again, allowing the consumer to proceed; the consumer will just have to wait a while. To implement this behavior, the runtime system first ensures that the lock on *a_buffer*'s processor is released (which was locked to allow the precondition to be evaluated); this allows values to be put in the buffer. The call is then reevaluated at a later point.

This means that the semantics of preconditions is reinterpreted: they are now treated as wait conditions, meaning that the execution of the body of the routine is delayed until they are satisfied. We can summarize this behavior in the *wait rule*:

> "A routine call with separate arguments will execute when all corresponding processors are available and the precondition is satisfied. The processors are held exclusively for the duration of the routine."

We complete the producer-consumer example by showing the code of the producer's main routine:

    *produce* (*a_buffer*: **separate** $BUFFER[\textbf{INTEGER}]$)
        **local**
            *value*: **INTEGER**
        **do**
            *value* := *random.produceValue*
            *a_buffer.put* (*value*)
        **end**

Since the buffer is unbounded, a wait condition is not necessary. It is however easily added and then makes the solution completely symmetric.

**Exercise A.10** What is the difference between the **require** clause in SCOOP and in Eiffel?

**Exercise A.11** Imagine a SCOOP routine has a precondition such as $n > 0$, that doesn't involve any separate targets. What do you think should happen in this case?

**Exercise A.12** You are to implement a controller for a device which can be accessed with the following interface:

>    **class** *DEVICE*
>        **feature**
>            *startup* **do** ... **end**
>            *shutdown* **do** ... **end**
>    **end**

There are also two sensors, one for heat and one for pressure, which can be used to monitor the device.

>    **class** *SENSOR*
>        **feature**
>            *value*: **INTEGER**
>            *device*: *DEVICE*
>    **end**

Write a class *CONTROLLER* in SCOOP that can poll the sensors concurrently to running the device. You should implement two routines: *run* starts the device and then monitors it with help of a routine *emergency_shutdown*, which shuts the device down if the heat sensor exceeds the value 70 or the pressure sensor the value 100.

**Exercise A.13** Name and explain three forms of synchronization used in SCOOP.

**Exercise A.14** Write down three possible outputs for the SCOOP program shown below:

```
class APPLICATION
create make
feature
  x: separate X
  y: separate Y
  z: Z

  make
    do
      create x; create y; create z
      print ("C")
      run1 (x)
      z.h
      run2 (y)
    end
  run1 (xx: separate X)
    do
      print ("A")
      xx.f
    end

  run2 (yy: separate Y)
    do
      yy.g (x)
      print ("L")
      yy.g (x)
    end
end
```

```
class X
feature
  n: INTEGER

  f
    do
      n := 1
      print ("K")
    end
end
```

```
class Z
feature
  h
    do
      print ("P")
    end
end
```

```
class Y
feature
  g (x: separate X)
    require
      x.n = 1
    do
      print ("Q")
    end
end
```

## A.5 Deadlock

While we have seen that locking is necessary for the proper synchronization of processes, it also introduces a new class of errors in concurrent programs: deadlocks. A *deadlock* is the situation where a group of processors blocks forever because each of the processors is waiting for resources which are held by another processor in the group. In SCOOP, the resources are the locks of the processors. As prescribed by the wait rule, a lock on processor $p$ is requested when executing a call to a routine with a separate argument handled by $p$; the lock is held for the duration of the routine.

As a minimal example, consider the following class:

```
class C
    creation
        make

    feature
        a : separate A
        b : separate A

        make (x : separate A, y : separate A)
            do
                a := x
                b := y
            end

        f do g (a) end
        g (x : separate A) do h (b) end
        h (y : separate A) do ... end
end
```

Now imagine that the following code is executed, where *c1* and *c2* are of type **separate** *C*, *a* and *b* are of type **separate** *A*, and *a* is handled by processor $p$, and *b* by processor $q$:

```
create c1.make (a, b)
create c2.make (b, a)
c1.f
c2.f
```

Since the arguments are switched in the initialization of *c1* and *c2*, a sequence of calls is possible that lets their handlers first acquire the locks to $p$ and $q$ respectively, such that they end up in a situation where each of them requires a lock held by the other handler.

Deadlocks are currently not automatically detected by SCOOP, and it is the programmers responsibility to make sure that programs are deadlock-free. An implementation of a scheme for preventing deadlocks is however underway, and is based on locking orders that prevent cyclical locking.

**Exercise A.15** Explain in detail how a deadlock can happen in the above example by describing a problematic sequence of calls and locks taken.

# Answers to the exercises

**Answer A.1** If all processes in a group are running at the same time, their execution is said to be *parallel*. If all processes of a group have started to execute but only one process is running at a time, their execution is said to be *interleaved*. We say that the execution of a group of processes is *concurrent* if it is either parallel or interleaved. □

**Answer A.2** A context switch is the exchange of one process's context (its program counter and CPU registers) with another process's context on a CPU. A context switch enables the sharing of a CPU by multiple processes. □

**Answer A.3** A process can be in one of three states: *running*, *ready*, and *blocked*. If a process is *running*, its instructions are currently executed on a processor; if a process is *ready*, it is waiting for the scheduler to be assigned to a CPU; if a process is *blocked*, it is currently waiting for an external event which will set its state to *ready*. □

**Answer A.4** An asynchronous feature call is executed on a different processor than the current one. This means it runs concurrently with other computations that are subsequently executed on the current processor. Ordinary sequential feature calls which are executed on the current processor are called synchronous. In SCOOP, a feature call $t.f$ where $t$ is separate (of some type **separate** $X$) will be executed asynchronously; if $t$'s type is non-separate, it will be executed synchronously. □

**Answer A.5** The computation takes at least 80 time units, as can be seen from the following sequence diagram.

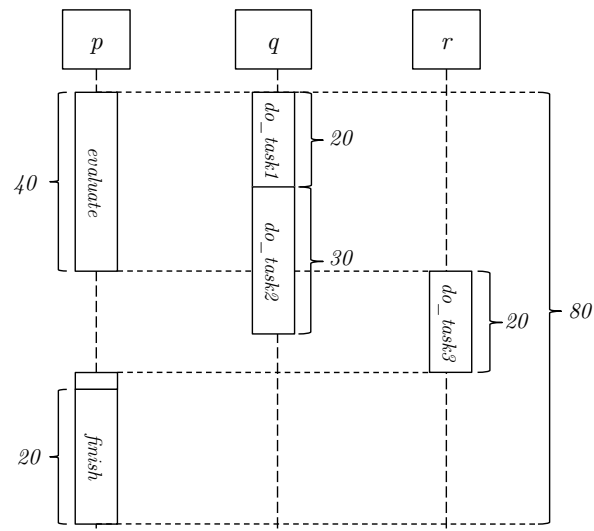

□

**Answer A.6** The following diagram depicts the object-processor associations and the references after execution of the program fragment.

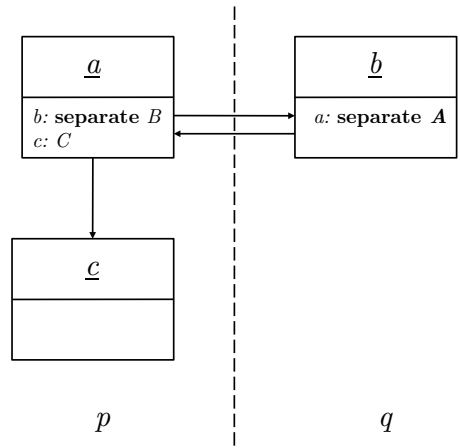

$\square$

**Answer A.7** A statement will be executed with wait-by-necessity semantics if it contains a query on a separate target. $\square$

**Answer A.8** A data race is the situation where the result of a concurrent computation depends on scheduling. Mutual exclusion is a form of synchronization to avoid the simultaneous use of a shared resource (such as a shared object) by multiple processes.

In SCOOP, an object can only be accessed by its handler, and this handler must be locked before it can be used to execute calls on the object. Mutual exclusion follows from the fact that only one processor can have a lock on another processor at any time. A lock on the handler of some object is taken by passing this object as an argument to the routine it is used in. SCOOP enforces this argument passing by the separate argument rule. $\square$

**Answer A.9** The class contains numerous violations of the separate argument rule. These violations are reported and fixed in the following code:

```
class MOTORBIKE
    feature
        engine: separate ENGINE
        front_wheel: separate WHEEL
        back_wheel: separate WHEEL
        display: DISPLAY

        initialize (e: separate ENGINE) -- Added separate argument
            do
                e.initialize -- Fixed: engine.initialize was incorrect as 'engine' is a
                    separate target, but not argument of the routine 'initialize'
                initialize_wheels(front_wheel, back_wheel)
                display.show ("Ready") -- This is correct: display is non-separate
            end

        initialize_wheels (f, b: separate WHEEL) -- Added separate arguments
            do
                display.show ("Initializing wheels ...")
                f.initialize -- Fixed
                b.initialize -- Fixed
            end
```

```
        switch_wheels
            local
                wheel: separate WHEEL -- Fixed: changed type from WHEEL to
                    separate WHEEL...
            do
                wheel := front_wheel -- ...otherwise this would violate the typing rule: a
                    separate source is assigned to a non-separate target
                front_wheel := back_wheel
                back_wheel := wheel
            end
    end
```

**Answer A.10** In ordinary Eiffel, a precondition that evaluates to false gives rise to an exception. In SCOOP no exception is thrown and instead the call is scheduled for reevaluation at a later point. □

**Answer A.11** A precondition that doesn't involve any separate targets will always evaluate to the same value, as the objects involved cannot be changed concurrently. If such a precondition evaluates to false, an exception is therefore thrown, just as in the sequential case. □

**Answer A.12** The controller can be implemented in the following manner:

```
class CONTROLLER
    create
        make

    feature
        device: DEVICE
        heat: separate SENSOR
        pressure: separate SENSOR

        make (d: DEVICE; h, p: separate SENSOR)
            do
                device := d
                heat := h
                pressure := p
            end

        run (d: DEVICE)
            do
                d.startup
                emergency_shutdown (d, heat, pressure)
            end

        emergency_shutdown (d: DEVICE; h, p: separate SENSOR)
            require
                h.value > 70 or p.value > 100
            do
```

*d.shutdown*
                **end**
        **end**

Note that the wait conditions on *emergency_shutdown* ensure that the shutdown is initiated only if the sensors exceed their threshold values. Observe that the separate argument rule is correctly abided by. □

**Answer A.13** There are three major forms of synchronization provided in SCOOP: mutual exclusion, condition synchronization, and wait-by-necessity. Mutual exclusion for object access is ensured by the separate argument rule. Condition synchronization (waiting until a certain condition is true) is provided via the reinterpretation of preconditions as wait conditions. Wait-by-necessity is provided for queries on separate targets and ensures that an object is only queried after all previous calls have been finished and causes the caller to wait for this. □

**Answer A.14** Three possible output sequences are:

- CAKPLQQ

- CAPKQLQ

- CAKPQLQ

In routine *make* "C" is always printed at the beginning. Then there are three non-separate calls, which will be worked off one after the other. In *run1*, "A" is always printed first, but then the call *xx.f* is separate, i.e. will execute asynchronously. Hence, "K" might be printed after "A", but also after "P" has been printed as a result of the call *z.h*. The call *yy.g (x)* proceeds only if $x.n = 1$ is true, i.e. after "K" has been printed. Since both calls to *yy.g (x)* are asynchronous, but *print("L")* is synchronous "L" may be printed before or after the first "Q", but must be printe before the second "Q". □

**Answer A.15** The following sequence of calls can happen. First *c1.f* is executed, leading to the call *g (a)*. Since *a* is a separate argument of routine *g*, its handler *p* gets locked. Then *c2.f* is executed, leading to the call *g (b)*, since the roles of *a* and *b* are switched in *c1* and *c2*; this means that *q* is locked. On processor *p*, the call *h (b)* is issued, thus requesting a lock on *q*; on processor *q*, the call *h (a)* is issued, thus requesting a lock on *p*: a deadlock has occurred as none of the processors can proceed any further. □

# B  SCOOP: Test

## B.1  Background information

In this part of the test, we would like to collect some information concerning your prior experience with concurrent programming.

What level of studies are you currently completing?
☐ Bachelor in Computer Science
☐ Master in Computer Science
☐ PhD in Computer Science
☐ Other:

Which semester are you currently completing?

### B.1.1  Prior experience with concurrency

Have you ever taken or are you currently taking a course other than Software Architecture that covers concurrent programming?
☐ Yes
☐ No
☐ No, but I studied it on my own (e.g. through online tutorials, books, ...)

If yes, what course was/is it and when did you take it? (Please provide details below.)

☐ Parallel programming @ ETH Zurich by T. Gross in Spring
☐ Concepts of concurrent computation @ ETH Zurich by B. Meyer in Spring
☐ Other courses:

How much of the self-study material on concurrency that you worked with today did you already know before?

| ☐ | ☐ | ☐ | ☐ | ☐ | ☐ | ☐ | ☐ | ☐ | ☐ | ☐ |
|---|---|---|---|---|---|---|---|---|---|---|
| none | 10% | 20% | 30% | 40% | 50% | 60% | 70% | 80% | 90% | all |

### B.1.2 Programming experience (sequential and concurrent)

| | (1: a novice ... 5: an expert) |
|---|---|
| Concerning your **general programming experience**, do you consider yourself... | 1 ☐ 2 ☐ 3 ☐ 4 ☐ 5 ☐ |
| Concerning your experience with **concurrent programming**, do you consider yourself... | 1 ☐ 2 ☐ 3 ☐ 4 ☐ 5 ☐ |
| Concerning your experience with the programming language **Eiffel**, do you consider yourself... | 1 ☐ 2 ☐ 3 ☐ 4 ☐ 5 ☐ |
| Concerning your experience with the programming language **Java**, do you consider yourself... | 1 ☐ 2 ☐ 3 ☐ 4 ☐ 5 ☐ |
| Concerning your experience with **Java Threads**, do you consider yourself... | 1 ☐ 2 ☐ 3 ☐ 4 ☐ 5 ☐ |
| Concerning your experience with **SCOOP**, do you consider yourself... | 1 ☐ 2 ☐ 3 ☐ 4 ☐ 5 ☐ |

### B.1.3 Self-study material

Where did you work through the self-study material?
  ☐ In the morning lecture    ☐ In the exercise class    ☐ At home

| | (1: strongly disagree ... 5: strongly agree) |
|---|---|
| The self-study material was easy to follow. | 1 ☐ 2 ☐ 3 ☐ 4 ☐ 5 ☐ |
| The self-study material provided enough **examples** to help me understand the subject. | 1 ☐ 2 ☐ 3 ☐ 4 ☐ 5 ☐ |
| The self-study material provided enough **exercises** to help me understand the subject. | 1 ☐ 2 ☐ 3 ☐ 4 ☐ 5 ☐ |
| I was able to complete the tutorial within 90 minutes. | 1 ☐ 2 ☐ 3 ☐ 4 ☐ 5 ☐ |
| The self-study material is a good alternative to the traditional lectures. | 1 ☐ 2 ☐ 3 ☐ 4 ☐ 5 ☐ |
| I feel confident that I will be able to solve the tasks in this test. | 1 ☐ 2 ☐ 3 ☐ 4 ☐ 5 ☐ |

Any comments on the self-study material:

## B.2 Sequential comprehension

Write down the output of the sequential Eiffel program shown below.

**Solution:** AFSTML

```
class APPLICATION
create
   make

feature
   a: A
   b: B
   c: C

   make
      do
         create a; create b;
         create c;
         print ("A")
         run (a, b)
         print ("L")
      end

   run (aa: A; bb: B)
      do
         aa.m(c)
         bb.n(c)

         if (aa.done_a and bb.
             done_b)
         then
            print ("M")
         end
      end

end
```

```
class A

feature

   done_a: BOOLEAN

   m (cc: C)
      do
         cc.f
         done_a := true
      end

end
```

```
class B

feature

   done_b: BOOLEAN

   n(cc: C)
      do
         cc.g
         done_b := true
         print ("T")
      end

end
```

```
class C

feature

   done_c: BOOLEAN

   f
      do
         print("F")
         done_c := false
      end

   g
      do
         print("S")
         done_c := true
      end

end
```

## B.3 General concurrency concepts

What is multiprocessing?

☐ Execution of multiple processes, within a single computer sharing a single processing unit.
☐ Execution of a single process on a single computer.
☐ Execution of a single process within multiple computers.
☐ Execution of multiple processes within a single computer sharing two or more processing units.

**Solution:** d

Which of the following state transitions is not possible in the status of a process?

☐ running → ready
☐ ready → blocked
☐ blocked → ready
☐ running → blocked

**Solution:** b

In the space below explain the terms data race and mutual exclusion.

**Solution:** See self-study material.

What is a deadlock?

**Solution:** See self-study material.

## B.4 Program comprehension

Write down three possible (non-deadlock) outputs for the SCOOP program shown below:

**Solution:** Some possible output sequences include: APTSFFTSML, ATSFPFTSML, ATSPFTSFML, APTTSSFFML, ATPTSFSFML, APTTSFFSML.

```
class APPLICATION
create
  make

feature
  a: separate A
  b: separate B
  c: separate C
  d: D

  make
    do
      create a; create b;
      create c; create d;
      print ("A")
      run (a, b)
      print ("L")
    end

  run (aa: separate A;
       bb: separate B)
    do
      aa.m (c)
      bb.n (c)
      aa.m (c)
      bb.n (c)
      d.foo

      if (aa.done_a and
          bb.done_b)
      then
        print ("M")
      end
    end
end
```

```
class A
feature
  done_a: BOOLEAN

  m (cc: separate C)
    require
      cc.done_c
    do
      cc.f
      done_a := true
    end
end
```

```
class B
feature
  done_b: BOOLEAN

  n(cc: separate C)
    require
      not cc.done_c
    do
      print ("T")
      cc.g
      done_b := true
    end
end
```

```
class C
feature
  done_c: BOOLEAN

  f
    do
      print("F")
      done_c := false
    end

  g
    do
      print("S")
      done_c := true
    end
end
```

```
class D
feature
  foo
    do
      print("P")
    end
end
```

## B.5 Program debugging

Identify errors (possibly compile-time) in the following SCOOP code segment. Justify your answers by providing on the next page the line number and a short explanation for every detected error. (The number of provided spaces does not necessarily correspond to the actual number of errors.)

```
 1  class A create make
 2  feature
 3    b: separate B
 4    c: C
 5
 6    make
 7      local
 8        b1: B;
 9      do
10        create b; create c;
11        b1 := g
12      end
13
14    f(b1: separate B): B
15      local
16        b2: B
17        c1: separate C
18      do
19        b2 := b
20        c1 := c
21        b.f
22        c.g
23        Result := b1.h
24      end
25
26    g: separate B
27      local
28        b1: B
29      do
30        h (b)
31        create b1
32        Result := b1
33      end
34
35    h(b1: B)
36      local
37        b2: separate B
38        c1: C
39        i: INTEGER
40      do
41        create b2
42        i := c.r
43        c1 := b2.h.r
44      end
45  end
```

```
46  class B
47
48  feature
49
50    r: C
51
52    h: B
53
54    f
55      do
56        create h
57        create r
58      end
59  end
```

```
60  class C
61
62  feature
63
64    r: INTEGER
65
66    g
67      do
68        r := 10
69      end
70  end
```

**Solution:** Some of the errors that could be mentioned include:

- Line 11: assignment of $b1 := g$ is not correct as $g$ returns a separate object.

- Line 19: assignment of $b2 := b$ is not correct as $b$ is separate.

- Line 21: $b.f$: routine needs to be wrapped. Violates the Separate Call rule: The target of a separate call must be a formal argument of the routine in which the call appears.

- Line 23 or Line 14: since $b1$ is separate, also $b1.h$ is separate with respect to the current object. However the result type is non-separate, which violates the typing rules.

- Line 30 or Line 35: $h(b)$ passes a separate entity as the actual parameter.

- Line 43: the right hand side of the assignment returns a separate entity, but $c1$ is non-separate.

## B.6 Program correctness

Consider a class *Data* with two integer fields $x$ and $y$, both of which are initialized to 0. Two classes *C0* and *C1* share an object *data* of type *Data*. Class *C0* implements the following behavior, which is repeated continuously: if both values *data.x* and *data.y* are set to 1, it sets both values to 0; otherwise it waits until both values are 1. Conversely, class *C1* implements the following behavior, which is also repeated continuously: if both values *data.x* and *data.y* are set to 0, it sets both values to 1; otherwise it waits until both values are 0. The following condition must always hold when *data* is accessed:

$$(data.x = 0 \land data.y = 0) \lor (data.x = 1 \land data.y = 1)$$

Write a concurrent program using SCOOP that implements the described functionality. Besides the mentioned classes *Data*, *C0*, and *C1*, your program needs to have a root class which ensures that the behaviors of *C0* and *C1* are executed on different processors.

**Solution:**

```
1  class
     APPLICATION

   create
5    make

   feature
     data: DATA
     c0: separate C0
10   c1: separate C1

     make
       do
         create data
15       create c0.make (data)
         create c1.make (data)

         run (c0, c1)
       end
20
     run (cc0: separate C0; cc1: separate C1)
       do
         cc0.run
         cc1.run
25     end

   end

   class DATA
30
   feature
     x, y: INTEGER
```

```
         set_x (v: INTEGER)
35         do
             x := v
           end

         set_y (v: INTEGER)
40         do
             y := v
           end

       end
45
       class C0

       create
         make
50
       feature
         data: separate DATA

         make (d: separate DATA)
55         do
             data := d
           end

         run
60         do
             from until False
             loop
               set_0 (data)
             end
65         end

         set_0 (d: separate DATA)
           require
             d.x = 1 and d.y = 1
70         do
             d.set_x (0)
             d.set_y (0)
           end
       end
75
       class C1

       create
         make
80
```

```
     feature
       data: separate DATA

       make (d: separate DATA)
85       do
           data := d
         end

       run
90       do
           from until False
           loop
             set_1 (data)
           end
95       end

       set_1 (d: separate DATA)
         require
           d.x = 0 and d.y = 0
100      do
           d.set_x (1)
           d.set_y (1)
         end
     end
```

## B.7 Feedback on the test

How much time did you spend on this test?

20' 30' 40' 50' 60' 70' 80' 90' 100' 110' 120'
☐ ☐ ☐ ☐ ☐ ☐ ☐ ☐ ☐ ☐ ☐

| | |
|---|---|
| The difficulty level of the test was... <br> (1: too easy, 2: easy, 3: just right, 4: difficult, 5: too difficult) | 1 2 3 4 5 <br> ☐ ☐ ☐ ☐ ☐ |
| I feel confident that I solved the tasks of this test correctly. <br> (1: strongly disagree ... 5: strongly agree) | 1 2 3 4 5 <br> ☐ ☐ ☐ ☐ ☐ |

Did you leave any questions of the test empty and if so, why?

Any comments on the test:

# C  Java Threads: Self-study material

Almost all computer systems on the market today have more than one CPU, typically in the form of a multi-core processor. The benefits of such systems are evident: the CPUs can share the workload amongst themselves by working on different instructions in parallel, making the overall system faster. This work sharing is unproblematic if the concurrently executing instructions are completely independent of each other. However, sometimes they need to access the same region of memory or other computing resources, which can lead to so-called *race conditions* where the result of a computation depends on the order of nondeterministic system events. Therefore concurrent processes have to be properly *synchronized*, i.e. programmed to wait for each other whenever necessary, and this calls for specialized programming techniques.

Today, you will learn about the background and techniques of *concurrent programming*. In particular, you will get to know the thread library approach to concurrent programming using the example of the *Java Threads* API. You might be familiar with Java Threads through other courses or previous self-study, in which case you should use this material to review your knowledge. At the end of this lesson, you will be able to

- explain the basics of concurrent execution of processes in modern operating systems, in particular multiprocessing and multitasking,

- understand some of the most important problems related to concurrent programming, in particular race conditions and deadlocks,

- distinguish between different types of process synchronization, in particular mutual exclusion and condition synchronization,

- understand how these types of synchronization are realized in Java Threads,

- program simple concurrent programs using Java Threads.

The lesson consists entirely of self-study material, which you should work through in the usual two lecture hours. You should have a study partner with whom you can discuss what you have learned. A the end of each study section there will be exercises that help you test your knowledge; solutions to the exercises can be found on the last pages of the document.

## C.1  Concurrent execution

This section introduces the notion of concurrency in the context of operating systems. This is also where the idea of concurrent computation has become relevant first, and as we all have to deal with operating systems on a daily basis, it also provides a good intuition for the problem. You may know some of this content already from an operating systems class, in which case you should see this as a review and check that you are familiar again with all the relevant terminology.

### C.1.1  Multiprocessing and multitasking

Up until a few years ago, building computers with multiple CPUs (Central Processing Units) was almost exclusively done for high-end systems or supercomputers. Nowadays, most end-user computers have more than one CPU in the form of a multi-core processor (for simplicity, we use the term CPU also to denote a processor core). In Figure 16 you see a system with two CPUs, each of which handles one process.

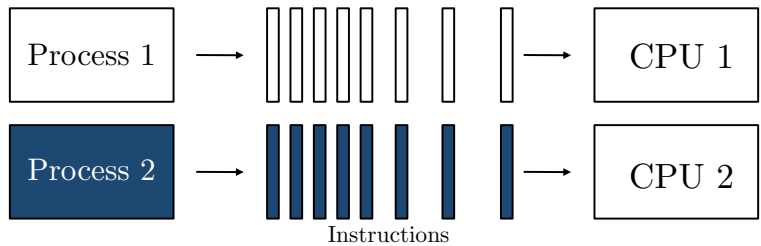

**Figure 16:** Multiprocessing: instructions are executed in parallel

The situation where more than one CPU is used in a single system is known as *multiprocessing*. The processes are said to execute *in parallel* as they are running at the same time.

However, also if you have a computer with a single CPU, you may still have the impression that programs run "in parallel". This is because the operating system implements *multitasking*, i.e. makes a single CPU appear to work at different tasks at once by switching quickly between them. In this case we say that the execution of processes is *interleaved* as only one process is running at a time. This situation is depicted in Figure 17. Of course, multitasking is also done on multiprocessing systems, where it makes sense as soon as the number of processes is larger than the number of available CPUs.

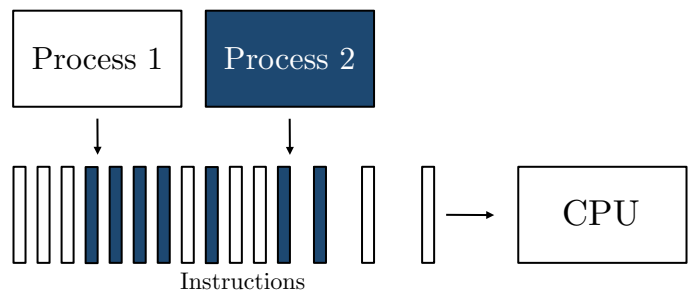

**Figure 17:** Multitasking: instructions are interleaved

Both multiprocessing and multitasking are examples of *concurrent execution*. In general, we say that the execution of processes is *concurrent* if they execute either truly in parallel or interleaved. To be able to reason about concurrent executions, one often takes the assumption that any parallel execution on real systems can be represented as an interleaved execution at a fine enough level of granularity, e.g. at the machine level. It will thus be helpful for you to picture any concurrent execution as the set of all its potential interleavings. In doing so, you will be able to detect any inconsistencies between different executions. We will come back to this point in Section C.3.1.

In the following section we will see how operating systems handle multitasking, and thus make things a bit more concrete.

### C.1.2 Operating system processes

Let's have a closer look at processes, a term which we have used informally before. You will probably be aware of the following terminology: a (sequential) *program* is merely a set of instructions; a *process* is an instance of a program that is being executed. The exact structure of a process may change from one operating system to the other; for our discussion it suffices to assume the following components:

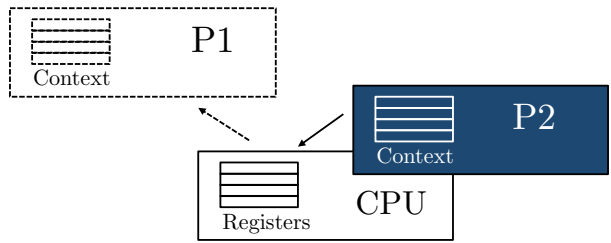

**Figure 18:** Context switch: process P1 is removed from the CPU and P2 is assigned to it

- *Process identifier*: the unique ID of a process.
- *Process state*: the current activity of a process.
- *Process context*: the program counter and the values of the CPU registers.
- *Memory*: program text, global data, stack, and heap.

As discussed in Section C.1.1, multiple processes can execute at the same time in modern operating systems. If the number of processes is greater than the number of available CPUs, processes need to be scheduled for execution on the CPUs. The operating system uses a special program called the *scheduler* that controls which processes are *running* on a CPU and which are *ready*, i.e. waiting until a CPU can be assigned to them. In general, a process can be in one of the following three states while it is in memory:

- *running*: the process's instructions are executed on a processor.
- *ready*: the process is ready to be executed, but is not currently assigned to a processor.
- *blocked*: the process is currently waiting for an event.

The swapping of process executions on a CPU by the scheduler is called a *context switch*. Assume a process P1 is in the state *running* and should be swapped with a process P2 which is currently *ready*, and consider Figure 18. The scheduler sets the state of P1 to *ready* and saves its context in memory. By doing so, the scheduler will be able to wake up the process at a later time, such that it can continue executing at the exact same point it had stopped. The scheduler can then use the context of P2 to set the CPU registers to the correct values for P2 to resume its execution. Finally, the scheduler sets P2's process state to *running*, thus completing the context switch.

From the state *running* a process can also get into the state *blocked*; this means that it is currently not ready to execute but waiting for some system event, e.g. for the completion of some prerequisite task by another process. When a process is *blocked* it cannot be selected by the scheduler for execution on a CPU. This can only happen after the required event triggers the state of the blocked process to be set to *ready* again.

**Exercise C.1** Explain the difference between parallel execution, interleaved execution, and concurrent execution.

**Exercise C.2** What is a context switch? Why is it needed?

**Exercise C.3** Explain the different states a process can be in at any particular time.

## C.2 Threads

Concurrency seems to be a great idea for running different sequential programs at the same time: using multitasking, all programs appear to run in parallel even on a system with a single CPU, making it more convenient for the user to switch between programs and have long-running tasks complete "in the background"; in the case of a multiprocessing system, the computing power of the additional CPUs speeds up the system overall.

Given these conveniences, it also seems to be a good idea to use concurrency not only for executing different sequential programs, but also within a single program. For example, if a program implements a certain time-intensive algorithm, we would hope that the program runs faster on a multiprocessing system if we can somehow *parallelize* it internally. A program which gives rise to multiple concurrent executions at runtime is called a *concurrent program*.

### C.2.1 The notion of a thread

Imagine the following method *compute* which implements a computation composed of two tasks:

```
void compute() {
    t1.doTask1();
    t2.doTask2();
}
```

Assume further that it takes $m$ time units to complete the call *t1.doTask1()* and $n$ time units to complete *t2.doTask2()*. If *compute()* is executed sequentially, we thus have to wait $m$ time units after the call *t1.doTask1()* before we can start on *t2.doTask2()*, and the overall computation will take $m + n$ time units, as shown in Figure 19.

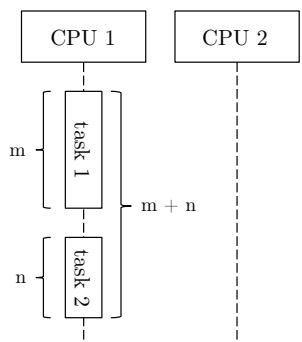

**Figure 19:** Sequential execution: the overall computation takes $m + n$ time units

If we have two CPUs, this seems rather a waste of time. What we would like to do instead is to execute *t1.doTask1()* on one of the CPUs and *t2.doTask2()* on the other CPU, such that the overall computation takes only $max(m, n)$ time units, as shown in Figure 20.

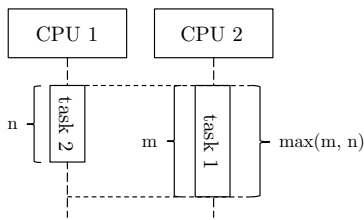

**Figure 20:** Parallel execution: the overall computation takes $max(m, n)$ time units

In order to be able to associate computation with different execution units, we introduce the notion of a *thread*. A thread can best be understood as a "lightweight process". This means that each thread has its own thread ID, program counter, CPU registers, and stack, and can thus support independent execution of instructions. However, threads are contained within processes, meaning that code and data memory sections and other resources belong to the containing process and are shared by all its threads. A process that has more than one such thread of control is called a multithreaded process. Threads can be assigned to physical CPUs via multitasking, just as operating systems processes are.

### C.2.2  Creating threads

A concurrent program gives rise to a multithreaded process at execution time, so the question is how we can create multiple threads in a programming language. In most programming languages, this is done via thread libraries, which provide the programmer with the API for managing threads. In this lecture we will use the the *Java Threads* API, however the concepts of libraries of other languages are quite similar.

Every Java program consists of at least one thread, which executes the *main*() method. In addition there is the possibility to define user threads; one way to do so is to inherit from the class *Thread* and to override its *run*() method. For example, the *run*() methods of the following two classes provide the implementations of *doTask1*() and *doTask2*() from above.

```
class Worker1 extends Thread {
    public void run() {
        // implement doTask1() here
    }
}
class Worker2 extends Thread {
    public void run() {
        // implement doTask2() here
    }
}
```

To create threads from these classes, one first creates a *Thread* object and then invokes the *start*() method on this object. This causes the *run*() method of the object to be executed in a new thread. Continuing the example, the following implementation of the method *compute*() creates two threads so that the two tasks from the example can be executed concurrently, and might finish in *max(m, n)* time units.

```
void compute() {
    Worker1 worker1 = new Thread1();
    Worker2 worker2 = new Thread2();
    worker1.start();
    worker2.start();
}
```

### C.2.3  Joining threads

Let's assume that the classes *Worker1* and *Worker2* from above are extended with the following method and attribute:

```
private int result;
```

```
public void getResult() {
    return result;
}
```

This allows the threads to save the final results of the computation in a variable, which can later be read out. For example, we might imagine that the results of the two tasks need to be combined in the *compute*() method:

```
return worker1.getResult() + worker2.getResult();
```

Clearly, we have to wait for both threads to be finished, before we can combine the results. This is done using the *join*() method which, when invoked on a supplier thread, causes the caller thread to wait until the supplier thread is terminated. For our example, this looks as follows:

```
int compute() {
    worker1.start();
    worker2.start();

    worker1.join();
    worker2.join();

    return worker1.getResult() + worker2.getResult();
}
```

Hence the main thread will first wait for *worker1* to finish, and then for *worker2*. Since we have to wait for both threads to finish, the order of the *join*() calls is arbitrary.

**Exercise C.4** Consider that the *run*() methods of threads *t1-t4* contain the following code

| t1: | worker1.doTask1(); worker2.doTask2(); |
| t2: | manager.evaluate(); |
| t3: | worker3.doTask3(); |
| t4: | manager.finish(); |

and that the following program fragment is executed in the *main*() thread:

```
t1.start();
t2.start();
t2.join();
t3.start();
t1.join();
t3.join();
result := worker2.getValue() + worker3.getValue();
t4.start();
```

Assume that the call *worker1.doTask1*() takes 20 time units until it returns, *worker2.doTask2*() 30 time units, *manager.evaluate*() 40 time units, *worker3.doTask3*() 20 time units, *manager.finish*() 20 time units; the queries *worker2.getValue*() and *worker3.getValue*() return immediately. What is the minimum time for execution of this program? Draw a sequence diagram to justify your answer.

## C.3 Mutual exclusion

Up until now, concurrency seems easy enough to handle. If we want to execute instructions concurrently with the rest of the program, we put these instructions in the *run*() method of a new class inheriting from *Thread*, create a corresponding object and call the *start()* method on it. At runtime, this gives rise to a new thread executing our instructions, and we are done. However, what happens if different threads interfere with each other, for example access and modify the same objects? We will see that this might change the results of computations in unexpected ways, and we thus have to avoid these situations by using a special type of synchronization called mutual exclusion. Luckily, Java has a simple mechanism for ensuring mutual exclusion.

### C.3.1 Race conditions

Consider the following class *Counter* which only has a single attribute *value*, and features to set and increment *value*.

```
class Counter {
    private int value = 0;

    public int getValue() {
        return value;
    }

    public void setValue(int someValue) {
        value = someValue;
    }

    public void increment() {
        value++;
    }
}
```

Now assume that an entity $x$ of type *Counter* is created and consider the following code:

```
x.setValue(0);
x.increment();
int i = x.getValue();
```

What is the value of $i$ at the end of this execution? Clearly, if this code was part of a sequential program, the value would be 1. In a concurrent setting where we have two or more threads, the value of $x$ can be read/modified by all of them. For example consider the following call executed concurrently by another thread:

```
x.setValue(2);
```

What is the value of $i$ now?

The answer is that, if these are the only threads running concurrently and $x$ references the same object in both cases, $i$ could have any of the values 1, 2, or 3. The reason for this is easily explained by looking at the thread interleavings that could be taken:

| x.setValue(2)         | x.setValue(0)         | x.setValue(0)         | x.setValue(0)         |
| x.setValue(0)         | x.setValue(2)         | x.increment()         | x.increment()         |
| x.increment()         | x.increment()         | x.setValue(2)         | int i = x.getValue()  |
| int i = x.getValue()  | int i = x.getValue()  | int i = x.getValue()  | x.setValue(2)         |
| --- | --- | --- | --- |
| i == 1, x.value == 1  | i == 3, x.value == 3  | i == 2, x.value == 2  | i == 1, x.value == 2  |

This is not really what we intended. The result of our computation has become arbitrary, and depends on the scheduling that determines a particular interleaving. Remember that we have no control over the scheduling.

The situation that the result of a concurrent execution is dependent on the nondeterministic scheduling is called a *race condition* or a *data race*. Data races are one of the most prominent problems in the domain of concurrent programming, and you can imagine that it gives rise to errors which can be quite hard to detect. For example, when you are running a program such as the above, say, 100 times, it might be that, because of a specific timing of events, you always obtain the values $i == 1$ and $x.value == 1$. But when you run the program for the 101st time, one of the other results arises. This means that such errors can stay hidden for a long time, and might never be detected during testing.

The question is now how to avoid data races. Java has a specific mechanism for this, which will be explained in the next section.

### C.3.2 Synchronized methods

To avoid data races we have to *synchronize* different computations such that they don't interfere with each other. Let's think about the main reason for the problem to occur. In the above example, two computations *shared* a resource, namely the object referenced by $x$. A part of a program that accesses a shared resource is called a *critical section*. The problem would not have occurred if, at any time, at most one computation would be in its critical section. The form of synchronization ensuring this property is called *mutual exclusion*.

Java provides the programmer with a simple way to ensure mutual exclusion. Each object in Java has a *mutex lock*, i.e. a lock that can be held by only one thread at a time. Thus to create a new lock, any object will do:

    *Object lock =* **new** *Object*();

A thread can acquire and release a lock using synchronized blocks:

    **synchronized** (*lock*) {
        *// critical section*
    }

When a thread reaches the start of the block, it tries to acquire the lock of the object referenced by *lock*. If the lock is held by another thread, the thread blocks until the lock is finally available. It will then acquire the lock and hold it until the control reaches the end of the block, where the lock is automatically released. As an example, imagine that the instructions from above are put into synchronized blocks, so in one thread we have

    **synchronized** (*lock*) {
        *x.setValue*(0);
        *x.increment*();
        **int** *i = x.getValue*();
    }

and in the other thread we have

    **synchronized** (*lock*) {
        *x.setValue*(2);
    }

where we make sure that the object referenced by *lock* is the same in both cases.

As explained above, since the same object referenced by *lock* acts as the lock in both synchronized blocks, the critical sections can be executed in mutual exclusion. This means that the state of the object referenced by $x$ can only be modified by one of the threads at a time, and hence upon completion of the first block we always obtain the result $i == 1$.

Besides having an explicit synchronized block, a method can also be decorated with the keyword **synchronized**. This has the same effect as enclosing the method body in a synchronized block where the current object **this** provides the lock, as shown in Figure 21.

```
synchronized type method(args) {

    // body

}
```

```
type method(args) {
    synchronized (this) {
        // body
    }
}
```

**Figure 21:** Correspondence between synchronized blocks and synchronized methods

**Exercise C.5** Explain the terms data race and mutual exclusion. How can we ensure mutual exclusion in Java Threads?

**Exercise C.6** Recall the *Counter* class from above, and imagine a class *SynchronizedCounter* which has all its methods declared as **synchronized**, but is otherwise identical to *Counter*. Find a simple example involving two threads where the result of the computation is nondeterministic when the methods from *Counter* are used, but not when the ones from *SynchronizedCounter* are used. Explain how these results come about.

## C.4 Condition synchronization

Protecting access to shared variables is not the only reason why a thread has to synchronize with other threads. For example, assume that a thread continuously takes data items out of a buffer to process them. Hence, the thread should only access the buffer if it holds at least one element; if it finds the buffer empty, it therefore needs to wait until another thread puts a data item in. Delaying a thread until a certain condition holds (as in this case, until the "buffer is not empty") is called *condition synchronization*. As you will see, in Java condition synchronization is enabled by the methods *wait()* and *notify()* which can be called on any synchronized object and allow a thread to release a previously acquired object lock and to notify other threads that a condition may have changed.

As an example of a problem that requires threads to use condition synchronization, we describe the so-called *producer-consumer problem*, which corresponds to issues found in many variations on concrete systems. Devices and programs such as keyboards, word processors and the like can be seen as *producers*: they produce data items such as characters or files to print. On the other hand the operating system and printers are the *consumers* of these data items. It has to be ensured that these different entities can communicate with each other appropriately so that for example no data items get lost.

On a more abstract level, we can describe the problem as follows. We consider two types of threads, both of which execute in an infinite loop:

- *Producer*: At each loop iteration, produces a data item for consumption by a consumer.
- *Consumer*: At each loop iteration, consumes a data item produced by a producer.

Producers and consumers communicate via a shared buffer implementing a queue; we assume that the buffer is unbounded, thus we only have to take care not to take out an item from an empty buffer, but are always able to insert new items. Instead of giving the full implementation we just assume to have a class *Buffer* to implement an unbounded queue:

*Buffer buffer =* **new** *Buffer*();

Producers append data items to the back of the queue using a method **void** *put*(**int** *item*), and consumers remove data items from the front using **int** *get*(); the number of items in a queue is queried by the method **int** *size*().

As part of the consumer behavior, we might for example want to implement the following method for consuming data items from the buffer:

```
public void consume() {
    int value;
    synchronized (buffer) {
        value = buffer.get(); // incorrect: buffer could be empty
    }
}
```

In this method we acquire a lock on the buffer using the synchronized block, and try to get an item from the buffer. The problem is that the buffer might be empty, i.e. *buffer.size*()== 0, which would result in a runtime error in this case. What we would like instead is that the thread waits before accessing the buffer, until the condition "buffer is not empty" is true, and then get the value from the buffer.

Waiting can be achieved in Java using the method *wait*(), which can be invoked on any object which is already locked, i.e. inside a synchronized block which has the object as lock; *wait*() then blocks the current thread (i.e. setting the thread state to *blocked*) and releases the lock. Continuing the example, we adapt the code as follows:

```
public void consume() throws InterruptedException {
    int value;
    synchronized (buffer) {
        while (buffer.size() == 0) {
            buffer.wait();
        }
        value = buffer.get();
    }
}
```

Note that the *wait*() call can throw an *InterruptedException* which we have to note in the method header (or otherwise in a try-catch block). Let's assume that the buffer is indeed found empty by the current thread. Thus the *wait*() call gets executed, the current process gets blocked and releases the lock on the object referenced by *buffer*.

Now the lock can be acquired by another thread, which might change the condition. To notify a waiting thread that the condition has changed, the thread can then execute the method *notify*() which unblocks one waiting thread (i.e. setting the thread state to *ready*), but doesn't yet release the lock. However, eventually the thread will release the lock by leaving the synchronized block, such that the unblocked thread can acquire it eventually and continue. Note that also the method *notify()* can only be called within a synchronized block which locks the object that it is called on.

This signaling step is part of the implementation of the method *produce()*:

```
public void produce() {
    int value = random.produceValue();
    synchronized (buffer) {
        buffer.put(value);
        buffer.notify();
    }
}
```

In this method the producer thread first creates a random integer value, locks the buffer and puts the value into the buffer. Then the thread uses *notify()* to signal a waiting consumer thread (if there is any) that the condition *buffer.size()== 0* is no longer true. Note however that the signaled consumer cannot take the truth of the condition for granted, as yet another interleaved consumer thread could have taken out the item before the signaled consumer was able to acquire the lock. This is why the checking of the condition *buffer.size()== 0* takes place within a while-loop, and not within an if-then-else: the unblocked consumer thread might need to block itself again.

It is important to note that a thread cannot know that a notification corresponds to the change of the condition it was interested in. For example, if we had a bounded buffer, we might also want to notify processes that the "buffer is not full". As *notify()* only unblocks a single process, we cannot be sure whether a process has been unblocked that waits for the condition "buffer is not full" or for "buffer is not empty". For this reason there is the method *notifyAll()* which unblocks all currently waiting processes. While we usually want to avoid using *notifyAll()* for efficiency reasons, and therefore we should use different lock objects corresponding to different conditions whenever possible.

**Exercise C.7** You are to implement a controller for a device which can be accessed with the following interface:

```
class Device {
    public void startup() { ... }
    public void shutdown() { ... }
}
```

There are also two sensors, one for heat and one for pressure, which can be used to monitor the device.

```
class Sensor extends Thread {
    Device device;
    private int value;

    public Sensor(Device d) {
        device = d;
    }

    public int getValue() {
        return value;
    }
```

```
        public void updateValue() { ... }

        public void run() { ... }
    }
```

Write a class *Controller* in Java Threads that can poll the sensors concurrently to running the device. You should implement its *run()* method such that it starts the device and then monitors it by waiting for and examining any new sensor values. The controller shuts down the device if the heat sensor exceeds the value 70 or the pressure sensor the value 100. Also complete the *run()* method in the class *Sensor* which calls *updateValue()* continuously and signals the controller if its value has changed.

**Exercise C.8** What is the difference between *notify()* and *notifyAll()*? When is it safe to substitute one with the other?

**Exercise C.9** Name and explain three forms of synchronization used in Java Threads.

**Exercise C.10** Write down three possible outputs for the Java Threads program shown below:

```java
public class Application
extends Thread {
  public static X x;
  public static Y y;
  public static Z z;

  public void run() {
    z = new Z(); x = new X(z);
    y = new Y(z);
    System.out.print("C");
    execute1();
    z.h();
    execute2();
  }

  public void execute1() {
    System.out.print("A");
    x.start();
  }

  public void execute2() {
    y.start();
    System.out.print("L");
  }
}

public class Root {
  public static void main(String[] args) {
    Application app = new Application();
    app.start();
  }
}
```

```java
class X
extends Thread {
  public Z z;

  public X(Z zz) {
    z = zz;
    z.n = 0;
  }

  public void run() {
    synchronized (z) {
      z.n = 1;
      z.notify();
      System.out.print("K");
    }
  }
}

class Z {
  public int n;
  public void h() {
    System.out.print("P");
  }
}
```

```java
class Y
extends Thread {
  public Z z;

  public Y(Z zz) {
    z = zz;
  }

  public void run() {
    System.out.print("J");
    synchronized (z) {
      while (z.n == 0) {
        try {
          z.wait();
        } catch
          (InterruptedException e) {};
      }
      System.out.print("Q");
    }
  }
}
```

## C.5 Deadlock

While we have seen that locking is necessary for the proper synchronization of processes, it also introduces a new class of errors in concurrent programs: deadlocks. A *deadlock* is the situation where a group of processors blocks forever because each of the processors is waiting for resources which are held by another processor in the group. In thread libraries, a common class of resources are mutex locks. As explained in Section C.3, locks are requested using synchronized blocks, and held for the duration of the method.

As a minimal example, consider the following class:

```
public class C extends Thread {
    private Object a;
    private Object b;

    public C(Object x, Object y) {
        a = x;
        b = y;
    }

    public void run() {
        synchronized (a) {
            synchronized (b) {
                ...
            }
        }
    }
}
```

Now imagine that the following code is executed, where *a1* and *b1* are of type *Object*:

```
C t1 = new C(a1, b1);
C t2 = new C(b1, a1);
t1.start();
t2.start();
```

Since the arguments are switched in the initialization of *t1* and *t2*, a sequence of calls is possible that lets the threads first acquire the locks to *a1* and *b1*, respectively, such that they end up in a situation where each of them requires a lock held by the other handler.

Note that there is no built-in mechanism of Java Threads that prevents deadlocks from happening, and it is the programmers responsibility to make sure that programs are deadlock-free.

**Exercise C.11** Explain in detail how a deadlock can happen in the above example by describing a problematic interleaving and the locks taken.

# Answers to the exercises

**Answer C.1** If all processes in a group are running at the same time, their execution is said to be *parallel*. If all processes of a group have started to execute but only one process is running at a time, their execution is said to be *interleaved*. We say that the execution of a group of processes is *concurrent* if it is either parallel or interleaved. □

**Answer C.2** A context switch is the exchange of one process's context (its program counter and CPU registers) with another process's context on a CPU. A context switch enables the sharing of a CPU by multiple processes. □

**Answer C.3** A process can be in one of three states: *running*, *ready*, and *blocked*. If a process is *running*, its instructions are currently executed on a processor; if a process is *ready*, it is waiting for the scheduler to be assigned to a CPU; if a process is *blocked*, it is currently waiting for an event which will set its state to *ready*. □

**Answer C.4** The computation takes at least 80 time units, as can be seen from the following sequence diagram.

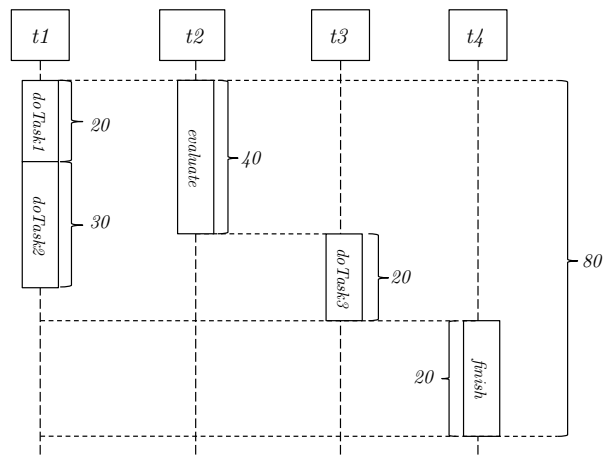

□

**Answer C.5** A data race is the situation where the result of a concurrent computation depends on scheduling. Mutual exclusion is a form of synchronization to avoid the simultaneous use of a shared resource (such as a shared object) by multiple processes.

Java Threads allows to protect critical sections by **synchronized** blocks. Two blocks guarded by the same lock object are guaranteed to execute in mutual exclusion. To protect a shared variable of an object, a common pattern is to declare the variable private in the class and declare all public methods accessing the shared variable **synchronized**. Then all method bodies will be executed with mutual exclusion (having the common lock object **this**), and the shared variable is protected against data races as it can only be accessed through these methods. □

**Answer C.6** The following simple example proves the point:

    *t1*:        *increment*();
    *t2*:        *increment*();

If we assume that the value of the counter is 0 at first, the value after both threads have finished will always be 2 in the case of the *SynchronizedCounter* methods, and 1 or 2 if *Counter* methods are used. The reason for this is that at the byte code level, the increment instruction consists of the following steps

```
        temp = value;
        temp = temp + 1;
        value = temp;
```

which can be interleaved in the case of *Counter*. In *SynchronizedCounter* this cannot happen as the **synchronized** block ensures that this group of instructions is executed atomically. □

**Answer C.7** The controller and the sensor can be implemented in the following manner, together with an appropriate root class:

```java
class Root {
  public static void main(String[] args) {
      Device d = new Device();
      Sensor h = new Sensor(d);
      Sensor p = new Sensor(d);
      Controller c = new Controller(d,h,p);
      h.start();
      p.start();
      c.start();
  }
}

class Controller extends Thread {
  Device device;
  Sensor heat;
  Sensor pressure;

  public Controller(Device d, Sensor h, Sensor p) {
    device = d;
    heat = h;
    pressure = p;
  }

  public void run() {
    device.startup();
    synchronized (device) {
      while (heat.getValue() <= 70 && pressure.getValue() <= 100) {
        try { device.wait(); } catch (InterruptedException e) {}
      }
    }
    device.shutdown();
  }
}

class Sensor extends Thread {
  Device device;
  private int value;

  public Sensor(Device d) {
    device = d;
```

```
      }

      public int getValue() {
        return value;
      }

      public void updateValue() { ... }

      public void run() {
        while (true) {
          synchronized (device) {
            int oldValue = value;
            updateValue();
            if (value != oldValue) {
              device.notify();
            }
          }
        }
      }
    }

    class Device {
      public void startup() { ... }
      public void shutdown() { ... }
    }
```

Note that condition synchronization is used to check on the emergency shutdown conditions: whenever the sensor obtains a new value, it will signal the controller, upon which the controller rechecks the condition and blocks itself if it is not yet fulfilled. □

**Answer C.8** The call *notify*() wakes up exactly one waiting thread, the call *notifyAll*() wakes up all waiting threads. A typical pattern for condition synchronization is

```
    while (!condition) {
        lock.wait();
    }
```

In this case it is safe to substitute any calls to *notify*() with *notifyAll*(), although this is inefficient: many threads which have been unblocked will just have to block themselves again as they find the condition invalidated once they get there.

On the other hand, *notifyAll*() typically cannot be substituted by *notify*() without semantic changes in other parts of the program. One reason is that threads can be blocked on various conditions within a **synchronized** block and that notify-calls cannot distinguish between them. The call *notifyAll*() will force all blocked processes to recheck their conditions, allowing at least one to proceed. The call *notify*() will only unblock one arbitrary process, whose condition might still be false; in this case a deadlock can happen, causing processes to wait for an event which will never manifest itself (see Section C.5). □

**Answer C.9** There are three major forms of synchronization provided in Java Threads: mutual exclusion, condition synchronization, and thread join. Mutual exclusion for object access can be ensured by the use of **synchronized** blocks. Condition synchronization (waiting until a certain condition is true) is provided via the methods *wait*() and *notify*() that can be called on synchronized objects at the point where the condition is known to be false or true, respectively. Joining threads is provided by the method *join*() which, called on a thread, causes the caller to wait until the thread has completed execution. □

**Answer C.10** Three possible output sequences are:

- CAPLKJQ
- CAPLJKQ
- CAKPLJQ

In method *run* of class *Application* "C" is always printed at the beginning. Upon calling *execute1*(), "A" is printed next and a new thread is started, which will run concurrently to the application thread. Hence "K" can be printed before or after "P" (which results from the call $z.h()$). "J" will be printed at the start of the thread $y$, after "P". Because of the waiting on the condition $z.n == 0$, "Q" will always be printed after "K". □

**Answer C.11** The following sequence of events can happen. First thread *t1* is executing, and obtains *a*'s lock at the start of its synchronized block. After a context switch, thread *t2* is executing and obtains *b*'s lock, since the roles of *a* and *b* are switched in *t1* and *t2*. After this *t1* will request *b*'s lock (currently held by *t1*), while *t2* requests *a*'s lock (currently held by *t2*): a deadlock has occurred as none of the threads can proceed any further. □

# D   Java Threads: Test

## D.1   Background information

In this part of the test, we would like to collect some information concerning your prior experience with concurrent programming.

What level of studies are you currently completing?
☐ Bachelor in Computer Science
☐ Master in Computer Science
☐ PhD in Computer Science
☐ Other:

Which semester are you currently completing?

### D.1.1   Prior experience with concurrency

Have you ever taken or are you currently taking a course other than Software Architecture that covers concurrent programming?
☐ Yes
☐ No
☐ No, but I studied it on my own (e.g. through online tutorials, books, ...)

If yes, what course was/is it and when did you take it? (Please provide details below.)

☐ Parallel programming @ ETH Zurich by T. Gross in Spring
☐ Concepts of concurrent computation @ ETH Zurich by B. Meyer in Spring
☐ Other courses:

How much of the self-study material on concurrency that you worked with today did you already know before?

| ☐ | ☐ | ☐ | ☐ | ☐ | ☐ | ☐ | ☐ | ☐ | ☐ | ☐ |
|---|---|---|---|---|---|---|---|---|---|---|
| none | 10% | 20% | 30% | 40% | 50% | 60% | 70% | 80% | 90% | all |

### D.1.2 Programming experience (sequential and concurrent)

|  | (1: a novice ... 5: an expert) |
|---|---|
| Concerning your **general programming experience**, do you consider yourself... | 1 2 3 4 5 ☐ ☐ ☐ ☐ ☐ |
| Concerning your experience with **concurrent programming**, do you consider yourself... | 1 2 3 4 5 ☐ ☐ ☐ ☐ ☐ |
| Concerning your experience with the programming language **Eiffel**, do you consider yourself... | 1 2 3 4 5 ☐ ☐ ☐ ☐ ☐ |
| Concerning your experience with the programming language **Java**, do you consider yourself... | 1 2 3 4 5 ☐ ☐ ☐ ☐ ☐ |
| Concerning your experience with **Java Threads**, do you consider yourself... | 1 2 3 4 5 ☐ ☐ ☐ ☐ ☐ |
| Concerning your experience with **SCOOP**, do you consider yourself... | 1 2 3 4 5 ☐ ☐ ☐ ☐ ☐ |

### D.1.3 Self-study material

Where did you work through the self-study material?
  ☐ In the morning lecture   ☐ In the exercise class   ☐ At home

|  | (1: strongly disagree ... 5: strongly agree) |
|---|---|
| The self-study material was easy to follow. | 1 2 3 4 5 ☐ ☐ ☐ ☐ ☐ |
| The self-study material provided enough **examples** to help me understand the subject. | 1 2 3 4 5 ☐ ☐ ☐ ☐ ☐ |
| The self-study material provided enough **exercises** to help me understand the subject. | 1 2 3 4 5 ☐ ☐ ☐ ☐ ☐ |
| I was able to complete the tutorial within 90 minutes. | 1 2 3 4 5 ☐ ☐ ☐ ☐ ☐ |
| The self-study material is a good alternative to the traditional lectures. | 1 2 3 4 5 ☐ ☐ ☐ ☐ ☐ |
| I feel confident that I will be able to solve the tasks in this test. | 1 2 3 4 5 ☐ ☐ ☐ ☐ ☐ |

Any comments on the self-study material:

## D.2 Sequential comprehension

Write down the output of the sequential Java program shown below.

**Solution:** AFSTML

| | | |
|---|---|---|
| ```
public class Application {
  public static A a;
  public static B b;
  public static C c;

  public Application() {
    c = new C(); a = new A(c);
    b = new B(c);
    System.out.print("A");
    execute();
    System.out.print("L");
  }

  public void execute() {
    a.m();
    b.n();
    if (a.done && b.done)
      System.out.print("M");
  }
}
``` | ```
public class A {
  C c;
  boolean done;

  public A(C cc) {
    c = cc;
    done = false;
  }

  public void m() {
    c.f();
    done = true;
  }
}
``` | ```
public class B {
  C c;
  boolean done;

  public B(C cc) {
    c = cc;
    done = false;
  }

  public void n() {
    c.g();
    done = true;
    System.out.print("T");
  }
}
``` |
| ```
public class Root {
  public static void main(String
      [] args) {
    Application app = new
        Application();
  }
}
``` | ```
public class C {
  boolean done;

  public void f() {
    System.out.print("F");
    done = false;
  }

  public void g() {
    System.out.print("S");
    done = true;
  }
}
``` | |

## D.3 General concurrency concepts

What is multiprocessing?

☐ Execution of multiple processes, within a single computer sharing a single processing unit.
☐ Execution of a single process on a single computer.
☐ Execution of a single process within multiple computers.
☐ Execution of multiple processes within a single computer sharing two or more processing units.

**Solution:** d

Which of the following state transitions is not possible in the status of a process?

☐ running → ready
☐ ready → blocked
☐ blocked → ready
☐ running → blocked

**Solution:** b

In the space below explain the terms data race and mutual exclusion.

**Solution:** See self-study material.

What is a deadlock?

**Solution:** See self-study material.

## D.4 Program comprehension

Write down three possible (non-deadlock) outputs for the Java Threads program shown below:

**Solution:** Some possible output sequences include: APTSFFTSML, ATSFPFTSML, ATSPFTSFML, APTTSSFFML, ATPTSFSFML, APTTSFFSML.

```java
public class Application
    extends Thread {
  public static A a;
  public static B b;
  public static C c;
  public static D d;

  public void run() {
    c = new C(); a = new A(c);
    b = new B(c); d = new D();
    System.out.print("A");
    execute();
    System.out.print("L");
  }

  public void execute() {
    a.start();
    b.start();
    d.foo();
    try {
      a.join();
      b.join();
    } catch
      (InterruptedException e) {};
    if (a.done_a && b.done_b)
      System.out.print("M");
  }
}
```

```java
public class A extends Thread {
  C c;
  boolean done_a;

  public A(C cc) {
    c = cc;
  }

  public void run() {
    try {
      m();
      m();
    } catch
      (InterruptedException e) {};
  }

  public void m() throws
      InterruptedException {
    synchronized (c) {
      while (!c.done) c.wait();
    }
    synchronized (c) {
      c.f();
      c.notifyAll();
    }
    done_a = true;
  }
}
```

```java
public class B extends Thread {
  C c;
  boolean done_b;

  public B(C cc) {
    c = cc;
  }

  public void run() {
    try {
      n();
      n();
    } catch
      (InterruptedException e) {};
  }

  public void n() throws
      InterruptedException {
    synchronized (c) {
      while (c.done) c.wait();
    }
    synchronized (c) {
      c.g();
      c.notifyAll();
    }
    done_b = true;
    System.out.print("T");
  }
}
```

```java
public class Root {
  public static void
      main(String[] args) {
    Application app =
        new Application();
    app.start();
  }
}
```

```java
public class C {
  boolean done;

  public void f() {
    System.out.print("F");
    done = false;
  }

  public void g() {
    System.out.print("S");
    done = true;
  }
}
```

```java
public class D {
  public void foo() {
    System.out.print("P");
  }
}
```

## D.5 Program debugging

Identify errors (possibly compile-time) in the following Java Threads code segment. Justify your answers by providing on the next page the line number and a short explanation for every detected error. (The number of provided spaces does not necessarily correspond to the actual number of errors.)

```java
1  class A extends Thread {
2    static B b;
3    static boolean done = false;
4
5    public static void main(String[] args) {
6      A a = new A();
7      b = new B();
8      C c = new C(b);
9
10     try {
11       a.start();
12       b.start();
13       c.start();
14       while (!done) {
15         System.out.println(b.g());
16       }
17       b.notifyAll();
18       a.join();
19       c.start();
20     }
21     catch (InterruptedException e) {}
22   }
23
24   public void run() {
25     int i;
26     for (i = 0; i < 100; i++) {
27       b.f();
28     }
29     done = true;
30   }
31 }
```

```java
1  class B {
2    int k = 0;
3
4    public synchronized int g() {
5      notify();
6      return k;
7    }
8
9    public void f() {
10     synchronized {
11       wait();
12       k++;
13     }
14   }
15 }
```

```java
1  class C extends Thread {
2    private B b;
3
4    public C(B bb) {
5      b = bb;
6    }
7
8    public void h() {
9      synchronized (b) {
10       b.k--;
11       notify();
12     }
13   }
14
15   public void run() {
16     int i;
17     for (i = 0; i < 100; i++) {
18       h();
19     }
20   }
21 }
```

**Solution:** Some of the errors that could be mentioned include:

- Line 12: Method *start()* cannot be called on *b*, since class *B* does not inherit from *Thread*
- Line 17: calling *notifiedAll()* on non-synchronized *b*
- Line 19: trying to start a thread which has already completed
- Line 41: **synchronized** { instead of **synchronized** (...){
- Line 42: *InterruptedException* for *wait()* not caught
- Line 57: *notify()* on non-synchronized **this**

## D.6 Program correctness

Consider a class *Data* with two integer fields $x$ and $y$, both of which are initialized to 0. Two classes *C0* and *C1* share an object *data* of type *Data*. Class *C0* implements the following behavior, which is repeated continuously: if both values *data.x* and *data.y* are set to 1, it sets both values to 0; otherwise it waits until both values are 1. Conversely, class *C1* implements the following behavior, which is also repeated continuously: if both values *data.x* and *data.y* are set to 0, it sets both values to 1; otherwise it waits until both values are 0. The following condition must always hold when *data* is accessed:

$$(data.x = 0 \land data.y = 0) \lor (data.x = 1 \land data.y = 1)$$

Write a concurrent program using Java Threads that implements the described functionality. Besides the mentioned classes *Data*, *C0*, and *C1*, your program needs to have a root class which ensures that the behaviors of *C0* and *C1* are executed in different threads.

**Solution:**

```
1  public class Application {
     public static void main(String[] args) {
       Data data = new Data();
       C0 c0 = new C0(data);
5      C1 c1 = new C1(data);
       c0.start();
       c1.start();
     }
   }
10
   public class Data {
     int x;
     int y;

15   public void set_x(int v) {
       x = v;
     }

     public void set_y(int v) {
20     y = v;
     }
   }

   public class C0 extends Thread {
25   private Data data;

     public C0(Data d) {
       data = d;
     }
30
     public void set_0() {
       synchronized (data) {
```

```
          while (!(data.x == 1 && data.y == 1)) {
            try {
35            data.wait();
            }
            catch(InterruptedException e) {}
          }
          data.set_x(0);
40        data.set_y(0);
          data.notifyAll();
        }
      }

45    public void run() {
        while (true) {
          set_0();
        }
      }
50  }

    public class C1 extends Thread {
      private Data data;

55    public C1(Data d) {
        data = d;
      }

      public void set_1() {
60      synchronized (data) {
          while (!(data.x == 0 && data.y == 0)) {
            try {
              data.wait();
            }
65          catch(InterruptedException e) {}
          }
          data.set_x(1);
          data.set_y(1);
          data.notifyAll();
70      }
      }

      public void run() {
        while (true) {
75        set_1();
        }
      }
    }
```

### D.7 Feedback on the test

How much time did you spend on this test?

<pre>
       20' 30' 40' 50' 60' 70' 80' 90' 100' 110' 120'
       ☐  ☐  ☐  ☐  ☐  ☐  ☐  ☐  ☐   ☐   ☐
</pre>

| | |
|---|---|
| The difficulty level of the test was... <br> (1: too easy, 2: easy, 3: just right, 4: difficult, 5: too difficult) | 1 2 3 4 5 <br> ☐ ☐ ☐ ☐ ☐ |
| I feel confident that I solved the tasks of this test correctly. <br> (1: strongly disagree ... 5: strongly agree) | 1 2 3 4 5 <br> ☐ ☐ ☐ ☐ ☐ |

Did you leave any questions of the test empty and if so, why?

Any comments on the test: